%% file: main.tex
\documentclass[12pt,aps,preprint,nofootinbib,superscriptaddress,nobalancelastpage]{revtex4}

\usepackage{mathrsfs}
\usepackage{amsmath,amsthm,amssymb}
\usepackage{color}
\usepackage{graphicx}
\usepackage{verbatim}

\usepackage{hyperref}
\hypersetup{colorlinks,bookmarksopen,bookmarksnumbered,citecolor=blue,
linkcolor=blue,pdfstartview=FitH,urlcolor=blue}
\usepackage{xspace}

\newcommand{\be}{\begin{equation}}
\newcommand{\ee}{\end{equation}}
\newcommand{\bea}{\begin{eqnarray}}
\newcommand{\eea}{\end{eqnarray}}

\newcommand{\Ne}{\ensuremath{\mathrm{N_{e}}}\xspace}
\newcommand{\Enu}{\ensuremath{\mathrm{E_{\nu}}}\xspace}

\definecolor{cadmiumgreen}{rgb}{0.0, 0.65, 0.31}

\begin{document}
\preprint{PCCF RI 15-04}

\title{Neutrino oscillations at DUNE with improved energy reconstruction}

\author{Valentina De Romeri}
\email{valentina.deromeri@uam.es}
\affiliation{Departamento de F\'isica Te\'orica, Universidad Aut\'onoma de Madrid, Cantoblanco E-28049 Madrid, Spain}
\affiliation{Instituto de F\'isica Te\'orica UAM/CSIC,
 Calle Nicol\'as Cabrera 13-15, Cantoblanco E-28049 Madrid, Spain}
\author{Enrique Fernandez-Martinez}
\email{enrique.fernandez-martinez@uam.es}
\affiliation{Departamento de F\'isica Te\'orica, Universidad Aut\'onoma de Madrid, Cantoblanco E-28049 Madrid, Spain}
\affiliation{Instituto de F\'isica Te\'orica UAM/CSIC,
 Calle Nicol\'as Cabrera 13-15, Cantoblanco E-28049 Madrid, Spain}
\author{Michel Sorel}
\email{sorel@ific.uv.es}
\affiliation{Instituto de F\'isica Corpuscular (IFIC), CSIC and Universitat de Valencia
Calle Catedr\'atico Jos\'e Beltr\'an, 2, 46980 Paterna, Valencia, Spain}

\begin{abstract}
{We study the physics reach of the long-baseline oscillation analysis of the DUNE experiment when realistic simulations are used to estimate its neutrino energy reconstruction capabilities. Our studies indicate that significant improvements in energy resolution compared to what is customarily assumed are plausible. This improved energy resolution can increase the sensitivity to leptonic CP violation in two ways. On the one hand, the CP-violating term in the oscillation probability has a characteristic energy dependence that can be better reproduced. On the other hand, the second oscillation maximum, especially sensitive to $\delta_{CP}$, is better reconstructed. These effects lead to a significant improvement in the fraction of values of $\delta_{CP}$ for which a $5 \sigma$ discovery of leptonic CP-violation would be possible. The precision of the $\delta_{CP}$ measurement could also be greatly enhanced, with a reduction of the maximum uncertainties from $26^\circ$ to $18^\circ$ for a 300~MW$\cdot$kt$\cdot$yr exposure. We therefore believe that this potential gain in physics reach merits further investigations of the detector performance achievable in DUNE.}
\end{abstract}

\maketitle

%
\input{intro}

\input{energyreconstruction}

\input{setup}

\input{results} 
\input{conclusions}

\section*{Acknowledgements}
We thank Daniel Cherdack and Elizabeth Worcester for useful comments and suggestions. We thank the DUNE Collaboration for the GDML file describing the full geometry of one far detector module used in our energy reconstruction analysis. We thank Pilar Coloma and other authors of \cite{Ankowski:2015jya} for providing their energy resolution results in a form suitable for comparison with our own. The authors acknowledge support from the EU through grants H2020-MSCA-ITN-2015/674896-Elusives and H2020-MSCA-RISE-2015/690575-InvisiblesPlus. EFM also acknowledges support from the EU FP7 Marie Curie Actions CIG NeuProbes (PCIG11-GA-2012-321582) and the Spanish MINECO through the ``Ram\'on y Cajal'' programme (RYC2011-07710), the project FPA2012-31880 and through the Centro de Excelencia Severo Ochoa Program under grant SEV-2012-0249. This work was also advanced during the stay of EFM at the Aspen Center for Physics, which is supported by the National Science Foundation grant PHY-1066293. This stay was also supported by a grant from the Simons Foundation. MS also acknowledges support from the Spanish MINECO through the project FIS2014-53371-C4-1-R and through the Centro de Excelencia Severo Ochoa Program under grant SEV-2014-0398. VDR acknowledges support by the Spanish MINECO through the project FPA2012-31880. This work was done in the framework of a ``D\'efi InPhyNiTi'' project (N2P2M-SF).

\bibliographystyle{apsrev}
\bibliography{refs}
\end{document}

%% file: intro.tex
\section{Introduction}
\label{sec:intro}

The experimental evidence for neutrino flavor change from the neutrino oscillation phenomenon can be explained consistently through three family mixing encoded in the unitary Pontecorvo--Maki--Nakagawa--Sakata~(PMNS) matrix~\cite{Pontecorvo:1957cp,Pontecorvo:1957qd,Maki:1960ut,Maki:1962mu,Pontecorvo:1967fh} in perfect analogy to the CKM matrix in the quark sector. However, the actual measurements of its elements reveal a structure completely different to its quark counterpart, increasing the complexity and mystery of the Standard Model (SM) flavor puzzle. Indeed, far from the hierarchical structure characterized by very small mixings of the CKM, large mixing angles are required to describe lepton mixing through the PMNS matrix. While the ``solar'' mixing angle has been measured to be $\theta_{12} \sim 33.5^\circ$, the ``atmospheric'' mixing angle $\theta_{23}$ is compatible with maximal mixing~\cite{Gonzalez-Garcia:2014bfa}. The discovery of a non-zero and relatively large $\theta_{13}$ mixing angle, $\theta_{13} \sim 8.5^\circ$~\cite{An:2012eh,Ahn:2012nd,Abe:2012tg,Adamson:2011qu,Abe:2011sj} now opens the window to the precision era in the measurements of the PMNS elements and to explore the existence of leptonic CP violation at present and near future oscillation facilities. Furthermore, the Jarlskog invariant $J = 1/8 \cos \theta_{13} \sin 2 \theta_{13} \sin 2 \theta_{12}\sin 2 \theta_{23} \sin\delta_{CP}$, which encodes the amount of CP violation present in the mixing matrix, can be potentially as large as $\sim 0.035$ for maximally CP violating values of $\delta_{CP}$, three orders of magnitude larger than the value of its counterpart in the quark sector: $J_{\rm CKM} = (2.96^{+0.20}_{-0.16})\times 10^{-5}$~\cite{Agashe:2014kda}. Since it has been shown that, within the context of Standard Model Electroweak Baryogenesis, $J_{\rm CKM}$ is not large enough to account for the observed Baryon Asymmetry of the Universe~\cite{Gavela:1993ts,Gavela:1994dt}, the discovery of an additional source of CP violation (such as $\delta_{CP}$ in the PMNS matrix) would be a very welcome input to tackle this open problem of the SM.

At present, T2K~\cite{Abe:2013hdq} and NO$\nu$A~\cite{Adamson:2016tbq} are providing the first tantalizing hints for maximal leptonic CP violation $\delta_{CP} \sim -90^\circ$ when combined with reactor neutrino results. While the actual significance of these hints is still low~\cite{Capozzi:2016rtj,Gonzalez-Garcia:2014bfa,nufit2_1}, they at least lead to a consistent picture. Nevertheless, a new generation of neutrino oscillation facilities, characterized by intense beams and large detectors, will be required to definitely settle the issue. These facilities will also unveil other present unknowns, such as the octant of $\theta_{23}$, if it turns out not to be maximal, and the neutrino mass hierarchy. Both these measurements have deep consequences in our understanding of the flavor puzzle. Indeed, deviations of $\theta_{23}$ from maximality can be crucial to understand if the observed flavor structure stems from a deeper underlying symmetry structure. The ordering of the neutrino mass eigenstates will reveal whether the spectrum is similar to that of quarks (with the lighter states having a larger mixing with the lighter charged lepton: the electron) for a normal hierarchy (NH) or opposite in that sense, for inverted (IH). The hierarchy determination is also a crucial input to neutrinoless double beta decay searches probing the Majorana nature of the neutrino fields. If neutrinos do indeed turn out to be Majorana particles, this would imply violation of lepton number and could again hint at a neutrino explanation of the observed Baryon Asymmetry of the Universe via the leptogenesis~\cite{Fukugita:1986hr} mechanism.

In this work, we re-evaluate the physics potential to address these unknowns of the proposed DUNE~\cite{Acciarri:2015uup} experiment exploiting the prospective LBNF facility at Fermilab. In particular, we analyze the energy reconstruction capabilities of the proposed liquid Argon (LAr) far detector and, with them, compute the physics reach of the experiment on the mass hierarchy, the octant of $\theta_{23}$ and the discovery of leptonic CP violation. In Section~\ref{sec:energyreconstruction} we discuss the simulations performed to characterize the energy reconstruction capabilities of the DUNE far detector. 
 In Section~\ref{sec:setup} we describe the experimental setup considered. In Section~\ref{sec:results} we analyze the sensitivity of the facility to the different physical observables in view of the previous results. Finally in Section~\ref{sec:conc} we draw our conclusions and summarize our results.

%% file: energyreconstruction.tex
\section{Neutrino energy reconstruction}
\label{sec:energyreconstruction}

In a neutrino detector, the energy of an incoming neutrino undergoing a charged-current (CC) interaction can be reconstructed either using charged lepton kinematics only (kinematic or quasi-elastic method), or using the entire visible energy in the event (calorimetric method), see for example \cite{Sorel:2014rka,Ankowski:2015jya}. For neutrino beams extending into the multi-GeV energy regime, as in the DUNE case, inelastic interactions are dominant and the calorimetric method is usually preferred. The performance of calorimetric neutrino energy reconstruction has been studied in detail in the context of the MINOS \cite{Michael:2008bc} and NO$\nu$A \cite{Adamson:2016xxw} experiments. For LAr-TPC neutrino detectors, studies primarily based on simulations exist, see for example \cite{Ferrari:2000wu,Rubbia:2011ft,Stahl:2012exa,Sorel:2014rka,Ankowski:2015jya,Ankowski:2015kya,Acciarri:2015uup}. Ongoing and planned efforts relying on LAr-TPCs exposed to charged particle \cite{Cavanna:2014iqa,Agostino:2014qoa,Kutter:2022751} and neutron \cite{Berns:2013usa} test beams are expected to provide additional calibration data for energy reconstruction purposes in the near future. In general, the calorimetric energy reconstruction of neutrino interactions in a LAr-TPC is expected to be driven by event-by-event fluctuations in the effects listed in Tab.~\ref{tab:energyreconstruction} \cite{Sorel:2014rka,Stahl:2012exa}.
\begin{table}[tbp]
\caption{\label{tab:energyreconstruction}Effects affecting calorimetric neutrino energy reconstruction in LAr-TPCs.}
\centering
\begin{tabular}{|c|l|} \hline
Number & Description \\ \hline
1 & Nuclear effects in neutrino interactions  \\ \hline 
2 & Non-deposited energy carried away by neutrinos  \\ \hline 
3 & Particle (other than neutrino) leakage out of the active volume  \\ \hline 
4 & Quenching of LAr ionisation/excitation from nuclear fragments  \\ \hline 
5 & Electron-ion recombination  \\ \hline 
6 & Electron attachment along drift  \\ \hline 
7 & Electronic noise of the charge read-out system \\ \hline 
\end{tabular}
\end{table}

In this work, we adopt a simple approach to energy reconstruction, extending the studies of Ref.~\cite{Sorel:2014rka}. We use detailed simulations of neutrino interactions and LAr response in a DUNE far detector module full geometry. We predict the total yield \Ne of ionisation electrons per event produced in the LAr active volume, for neutrino CC interactions of all relevant neutrino flavors and energies. We use \Ne as our estimator for the incoming neutrino energy \Enu, and hence event-by-event fluctuations in \Ne to estimate the DUNE far detector energy resolution performance. In our approach, we do not attempt to identify final state particles or neutrino interaction types. The full event charge produced in the LAr active volume is simply summed together, regardless of the final state particles producing it, and regardless of the time delay or spatial separation from the neutrino interaction vertex. In other words, we fully account for effects 1 to 5 in Tab.~\ref{tab:energyreconstruction}. We neglect electron attachment along drift (item 6) and electronic noise (item 7) contributions. The impact of these two items on calorimetric neutrino energy reconstruction depends on the details of the LAr-TPC detector under consideration, particularly on the LAr purity and the electronic noise levels. Their impact also depends on the actual reconstruction algorithms used, such as the hit finding algorithm to identify the read-out elements (wires or strips) and time samples carrying non-zero charge information, and the charge attenuation correction as a function of drift length. In the following, we assume that these effects have a negligible effect on calorimetric energy resolution compared to the other sources in Tab.~\ref{tab:energyreconstruction}. This is a plausible assumption, if we consider the impact of LAr-TPC full reconstruction effects obtained from earlier studies with simpler (non-neutrino) event topologies. In particular, ICARUS inferred a resolution of $\Delta$E/E = 1.0\%/$\sqrt{\rm{E~(GeV)}}$+1.2\% for electromagnetic showers induced by 0.05--5~GeV single photons \cite{Ankowski:2008aa}. As we will see, contributions at this level would be negligible compared to the neutrino energy resolutions we obtain.

We note that our simple approach to calorimetric neutrino energy reconstruction could be refined, potentially yielding better energy resolutions than the ones presented here. Particle or event identification could be exploited for calorimetry, thanks to known differences in detector response for different particles or events. In addition, scintillation light information may provide an additional handle to improve calorimetric energy resolution, as discussed in \cite{Sorel:2014rka}.

In the following, we discuss our energy reconstruction results for CC neutrino interactions (Sec.~\ref{subsec:energyreconstruction_larsoft}), how our strategy and results compare with previous studies (Sec.~\ref{subsec:energyreconstruction_cdr}), and how we treat energy reconstruction for neutral-current (NC) neutrino interactions (Sec.~\ref{subsec:energyreconstruction_nc}).


\subsection{LArSoft studies for charged-current neutrino interactions}
\label{subsec:energyreconstruction_larsoft}

We use the LArSoft code (\cite{Church:2013hea}, version v04\_09\_00) for our energy reconstruction studies. LArSoft is the software framework for simulation and reconstruction adopted by all LAr neutrino experiments, DUNE included. In particular, we use LArSoft for the detector geometry description, the simulation of neutrino-argon interactions, the propagation of final state particles in the detector active volume and surroundings, and for the description of the resulting ionization yields. We discuss each one of these aspects below.

The LAr-TPC detector geometry description used in LArSoft corresponds to one of the four far detector modules for DUNE at the Sanford Underground Research Facility (SURF). This geometry, in GDML format, has been provided to us by the DUNE Collaboration. It is equivalent to the ``workspace'' GDML geometries provided in \cite{dunecdrconfigs}, with the exception that it describes the full, 17~kt total LAr mass, detector module as opposed to only part of it. The detector enclosure, cryostat, 300 TPCs envisaged within the detector module, and the anode/cathode plane assemblies, are simulated. We do not consider readout effects in our studies, and hence we make use of the geometry description that does not include LAr-TPC wires, as discussed in \cite{dunecdrconfigs}. The dimensions of the LAr active volume extend from -7.46~m to +7.46~m in the X direction, from -6.08~m to +6.08~m in the Y direction, and from 0 to +58.1~m in the Z direction. The coordinate system convention adopted is such that Z is the beam direction and pointing downstream of the detector, Y is the vertical direction and pointing up, and the TPC drift direction X is taken to form a right-handed coordinate system. 

Neutrino-argon interactions are generated uniformly in the full LAr active volume using LArSoft, which in turn relies on the GENIE event generator code, version 2.8.6 \cite{Andreopoulos:2009rq}. Our energy resolution studies assume mono-energetic neutrinos of $\nu_e$, $\bar{\nu}_e$, $\nu_{\mu}$, $\bar{\nu}_{\mu}$ flavors, stepping through the relevant 0.25--10~GeV neutrino energy range. GENIE accounts for a comprehensive list of quasi-elastic, resonant, deep inelastic and coherent CC interaction channels. Nuclear effects, affecting both the initial and the final state of the interaction, are also accounted for.

All particles present in the neutrino-argon interaction final state are then propagated in LAr using the GEANT4-based \cite{Agostinelli:2002hh} detector simulation in LArSoft. The simulation uses the QGSP\_BERT physics list \cite{Yarba:2012ih}, using a Bertini cascade model for hadrons of energy below $\sim$10~GeV. The list contains all standard electromagnetic and weak decay processes. Compared to the default LArSoft physics list, we disable the neutron tracking cut, as in \cite{Sorel:2014rka}, to track neutrons down to thermal energies. 

Charged particles deposit energy by ionizing and exciting argon atoms, ultimately producing charge and light signals to be detected at readout planes. The simulation of charge and scintillation light signals is performed via the NEST \cite{Szydagis:2011tk} model, as in \cite{Sorel:2014rka}. The model accounts for quenching of LAr ionisation/excitation from nuclear fragments because of dissipation in the form of heat, and for electron-ion recombination. For the latter, a modified Birks' formula \cite{Doke:1988dp} and a 0.5~kV/cm drift field are assumed. We have verified that our energy reconstruction results are nearly insensitive to the details of the ionisation model chosen, with similar results obtained using the LArSoft default model \cite{Church:2013hea} compared to the more detailed NEST model \cite{Szydagis:2011tk}.

For our energy reconstruction studies, we assume the following fiducial volume definition (numbers in meters): [-6.5,-0.5]$\cup$[+0.5,+6.5] in X, [-5.0,+5.0] in Y, [+1.0,+54.0] in Z, for a LAr fiducial mass per detector module of about 9~kt. This fiducial mass is similar to the 10~kt value assumed in the DUNE CDR \cite{Acciarri:2015uup}. The gap in the fiducial volume near X=0 is due to the anode plane assembly. Our fiducial volume selection has been chosen by requiring an approximately constant average deposited energy as a function of (X,Y,Z) for relatively high energy (6~GeV) $\nu_e$ CC events, where the coordinates indicate the neutrino interaction location. Interactions occurring outside the fiducial volume are characterized by a rapidly decreasing average deposited energy, because of increased particle leakage out of the LAr volume. In this outer region, the detector would therefore have diminished neutrino energy reconstruction capabilities. 

\begin{figure}[tbp]
\centering
\includegraphics[width=.49\textwidth]{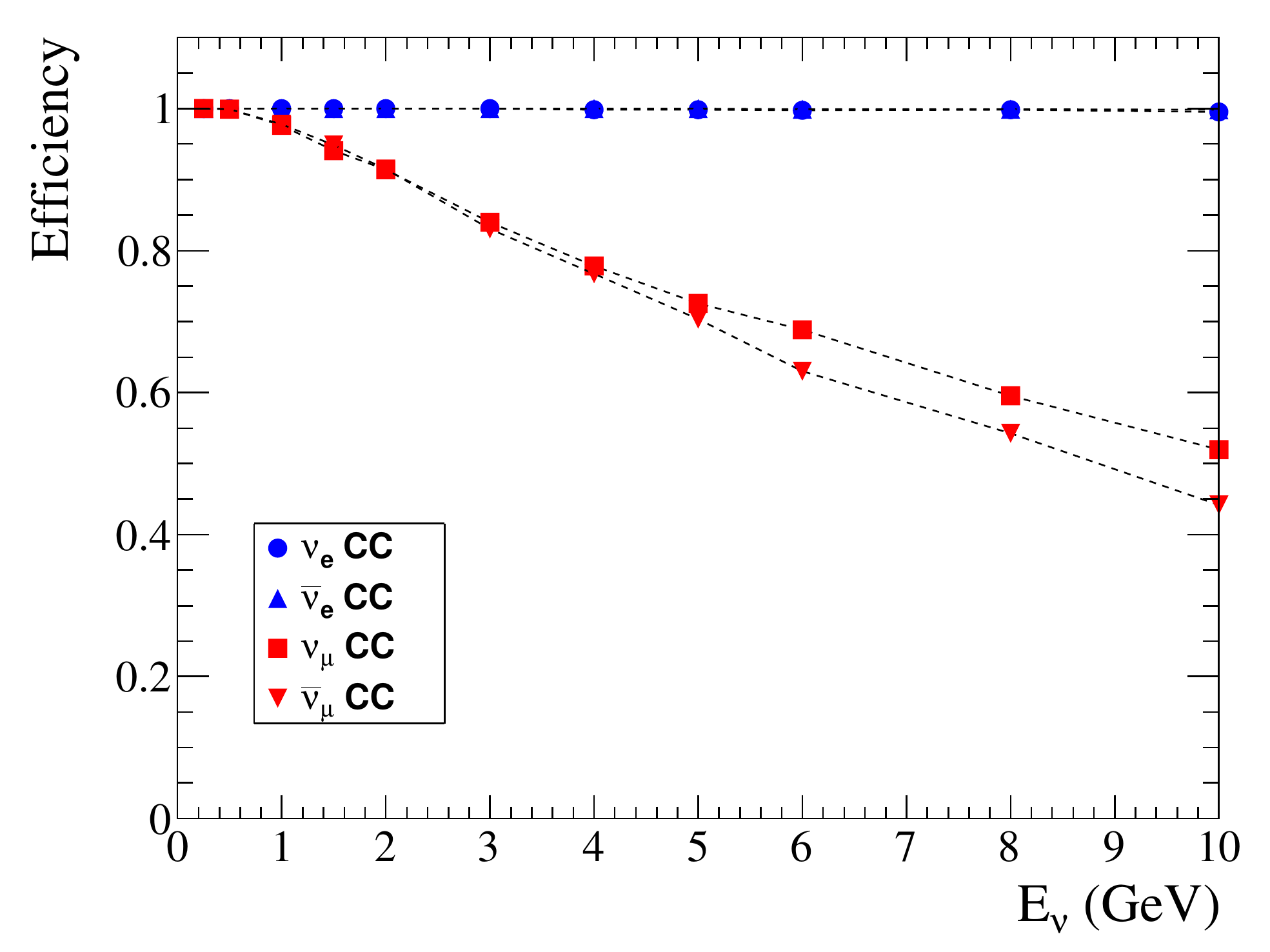}
\caption{\label{fig:ContainedLeptonEfficiency}Expected efficiency to fully contain the primary lepton within the LAr active volume, for $\nu_e$, $\bar{\nu}_e$, $\nu_{\mu}$ and $\bar{\nu}_{\mu}$ CC interactions in the LAr fiducial volume. The efficiency is shown as a function of neutrino energy.}
\end{figure}

In addition to this fiducial volume requirement, we only select neutrino CC interactions where the primary lepton (that is, the charged lepton produced at the neutrino CC interaction vertex) is fully contained within the LAr active volume. This is motivated by the desire of obtaining the best possible neutrino energy resolution. The efficiencies for primary lepton containment are shown in Fig.~\ref{fig:ContainedLeptonEfficiency}. While the lepton containment efficiency is 100\% for fiducial $\nu_e$ and $\bar{\nu}_e$ CC interactions at all relevant energies, the efficiency decreases with neutrino energy for $\nu_{\mu}$ and $\bar{\nu}_{\mu}$ CC interactions, as expected. However, even for 6~GeV $\nu_{\mu}$ and $\bar{\nu}_{\mu}$ CC, the efficiency is greater than 60\%.

\begin{figure}[tbp]
\centering
\includegraphics[width=.49\textwidth]{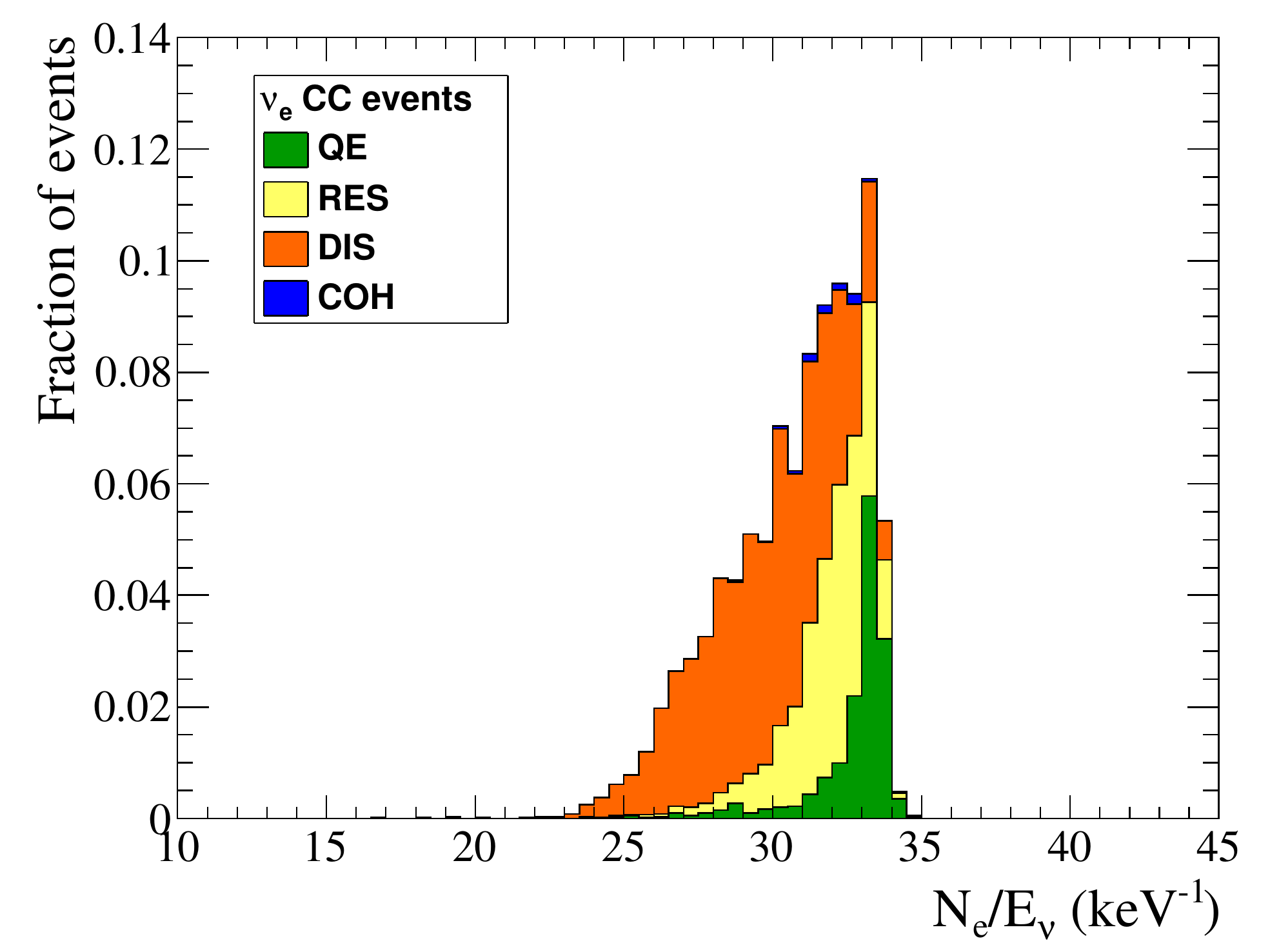} \hfill
\includegraphics[width=.49\textwidth]{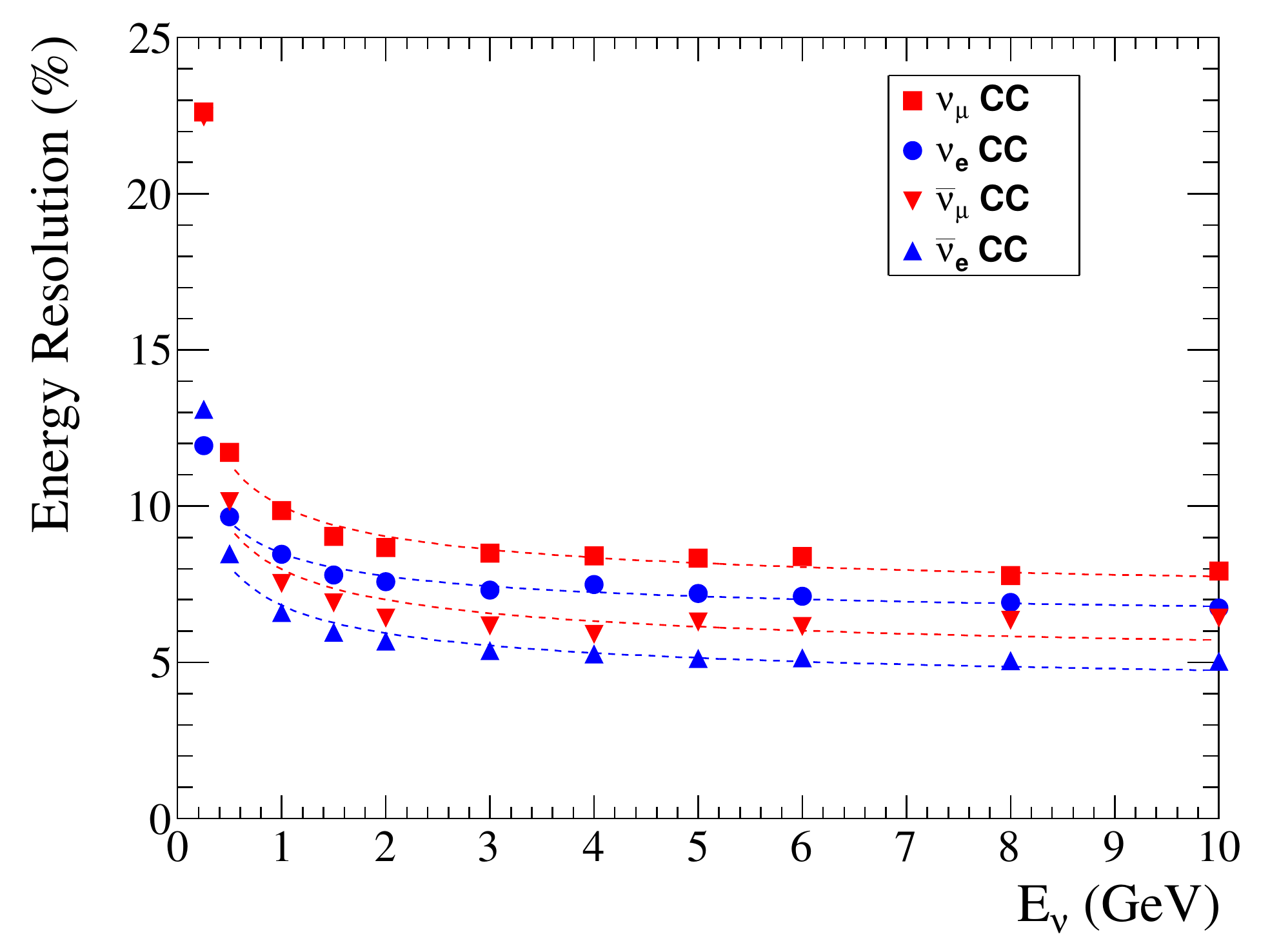}
\caption{\label{fig:EnergyReconstruction}Left panel: number of ionization electrons \Ne produced in the LAr active volume per incoming neutrino energy \Enu, for 4~GeV $\nu_e$ CC interactions throughout the LAr fiducial volume. The colors indicate different interaction types: quasi-elastic scattering (QE), resonant interactions (RES), deep inelastic scattering (DIS) and coherent pion production interactions (COH). Right panel: neutrino energy resolution expected from our LArSoft simulations and as a function of neutrino energy, for $\nu_{e}$, $\bar{\nu}_{e}$, $\nu_{\mu}$ and $\bar{\nu}_{\mu}$ CC interactions. The dashed lines correspond to the parametrizations given in the text.} 
\end{figure}

As mentioned above, we use the total ionization charge produced in the LAr active volume per neutrino interaction, \Ne, as our estimator for the incoming neutrino energy \Enu. The left panel of Fig.~\ref{fig:EnergyReconstruction} shows the \Ne/\Enu distribution for 4~GeV $\nu_e$ CC interactions in the DUNE fiducial volume as an example. On average, about 30 electron-ion pairs per keV of neutrino energy are produced. This number is significantly smaller than the 42~electrons/keV value expected from the ionization yield in LAr \cite{Miyajima:1974zz}. The difference is due to the detection effects mentioned above, particularly electron-ion recombination. These same effects (items 1--5 in Tab.~\ref{tab:energyreconstruction}) are also responsible for the event-by-event fluctuations in \Ne/\Enu. We estimate the neutrino energy resolution as the RMS width of \Ne/\Enu distributions such as the one in the left panel of Fig.~\ref{fig:EnergyReconstruction}.

The relative neutrino energy resolution for $\nu_e$, $\bar{\nu}_e$, $\nu_{\mu}$, $\bar{\nu}_{\mu}$ CC interactions and as a function of neutrino energy obtained from this procedure is shown in the right panel of Fig.~\ref{fig:EnergyReconstruction}. From the figure, energy resolutions of order 10\% or better are expected, with only a mild dependence on neutrino flavor or neutrino energy. For our physics studies, we parametrize the neutrino energy resolution via gaussian functions with flavor- and energy-dependent widths given by:
\begin{align}
\nu_e\ \mbox{CC:\hspace{1.cm}} & \Delta E_{\nu}/E_{\nu} = 0.060 + 0.025/\sqrt{E_{\nu}} \label{eq:nueccres} \\
\nu_{\mu}\ \mbox{CC:\hspace{1.cm}} & \Delta E_{\nu}/E_{\nu} = 0.067 + 0.033/\sqrt{E_{\nu}}  \label{eq:numuccres} \\
\bar{\nu}_e\ \mbox{CC:\hspace{1.cm}} & \Delta E_{\nu}/E_{\nu} = 0.038 + 0.030/\sqrt{E_{\nu}}  \label{eq:nuebarccres} \\
\bar{\nu}_{\mu}\ \mbox{CC:\hspace{1.cm}} & \Delta E_{\nu}/E_{\nu} = 0.047 + 0.033/\sqrt{E_{\nu}}  \label{eq:numubarccres} 
\end{align}
\noindent where the neutrino energy $E_{\nu}$ is expressed in GeV. The parametrizations in Eqs.~(\ref{eq:nueccres}--\ref{eq:numubarccres}), also shown in Fig.~\ref{fig:EnergyReconstruction} as dashed lines, are obtained by fitting our LArSoft energy resolution results in the 0.5--10~GeV energy range.

\begin{figure}[tbp]
\centering
\includegraphics[width=.49\textwidth]{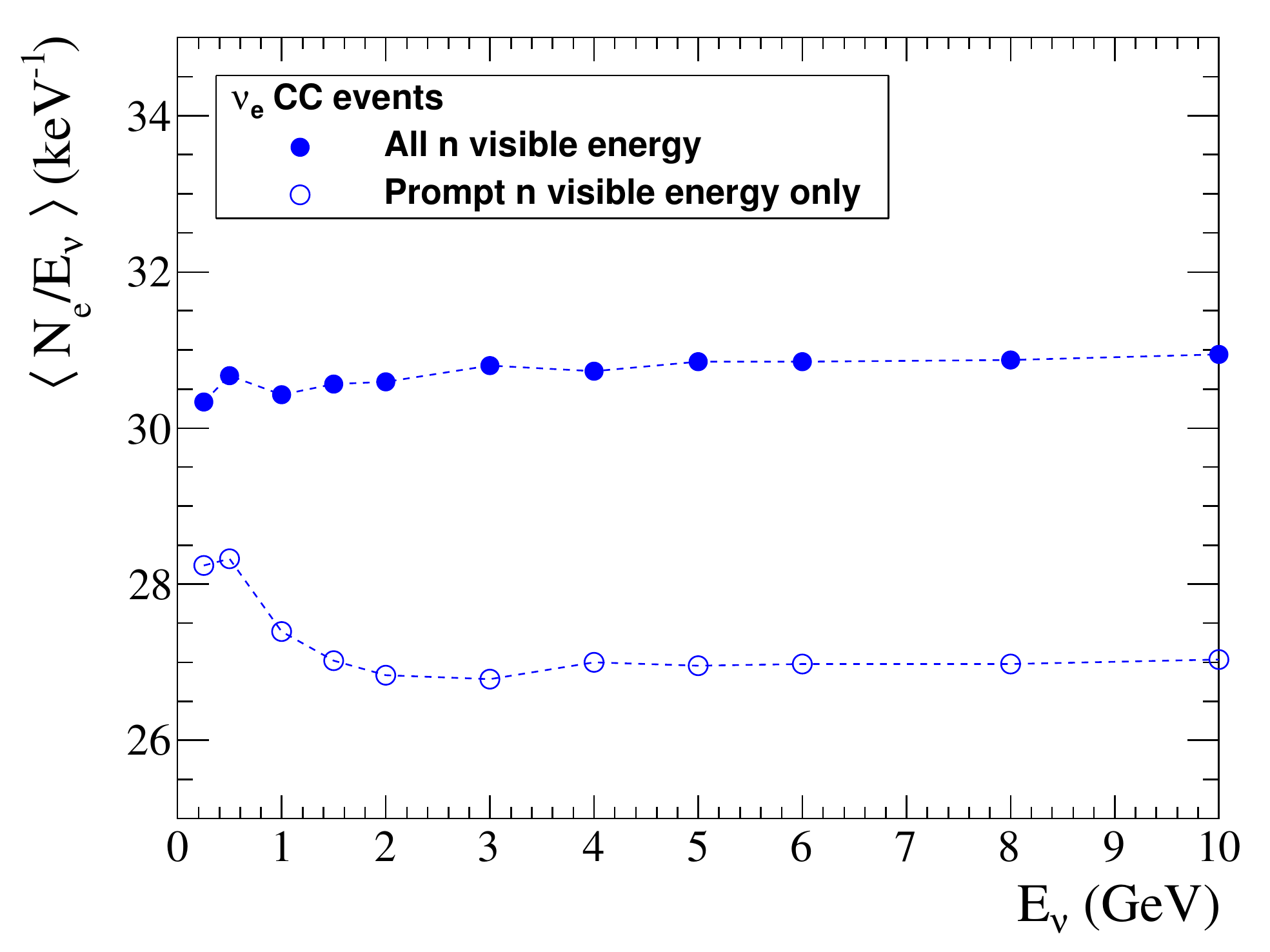} \hfill
\includegraphics[width=.49\textwidth]{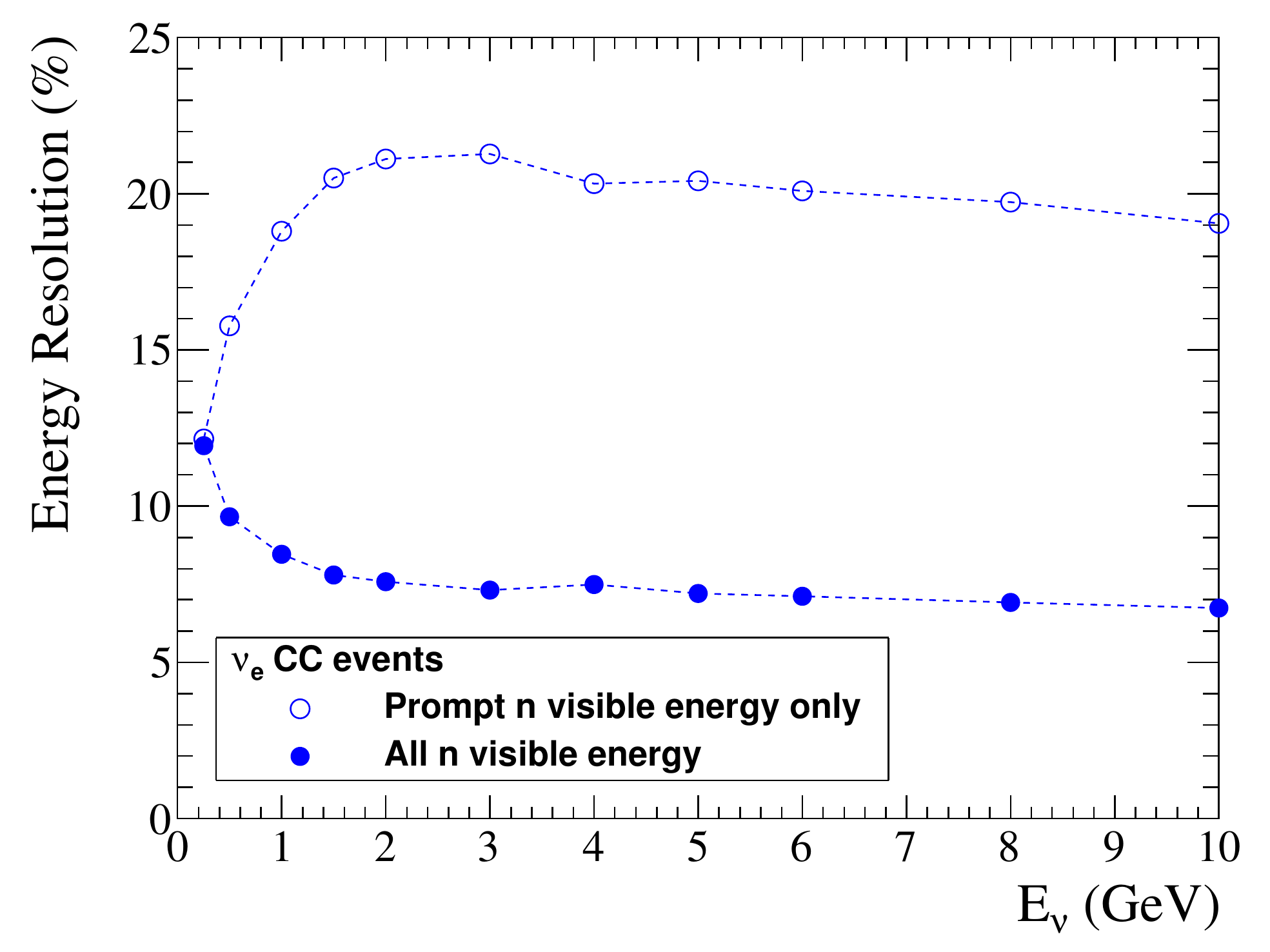}
\caption{\label{fig:EnergyReconstructionNeutronVariations}Average energy scale parameter $\langle \Ne/\Enu\rangle$ (left) and neutrino energy resolution (right) versus \Enu for fiducial $\nu_e$ CC events, for two different assumptions concerning DUNE neutron detection capabilities (see text for details).} 
\end{figure}

The importance of DUNE's neutron detection capabilities for calorimetric neutrino energy reconstruction is illustrated in Fig.~\ref{fig:EnergyReconstructionNeutronVariations}. The left panel of Fig.~\ref{fig:EnergyReconstructionNeutronVariations} shows average ionization yields per unit neutrino energy, $\langle \Ne/\Enu\rangle$, as a function of energy and for two different neutron detection assumptions. In both cases, only the ionization signals created as a result of neutron interactions in LAr are considered as neutron visible energy. The higher ionization yield assumes that all neutron-induced ionization in the LAr active volume can be detected, regardless of the delay or distance from the neutrino interaction time or vertex. The lower yield assumes that only the prompt neutron visible energy is detected, where prompt is defined as occurring within 10~$\mu$s from the neutrino interaction. For both assumptions, we find that $\langle \Ne/\Enu\rangle$ is stable to within $\simeq$2\% over the entire neutrino energy range of relevance for DUNE. Our simulations thus confirm that the total charge produced in the LAr is a good neutrino energy estimator. A larger effect is seen on the overall energy scale, with the less efficient neutron detection scenario collecting approximately 15\% less charge. While our work focuses on neutrino energy resolution, the left panel of Fig.~\ref{fig:EnergyReconstructionNeutronVariations} also gives us some indications on the scale of possible neutrino energy reconstruction biases in DUNE. Energy-scale uncertainties can affect long-baseline oscillation sensitivities \cite{Acciarri:2015uup} and, if not properly accounted for, can even introduce biases in the extracted oscillation parameters such as $\delta_{CP}$ \cite{Ankowski:2015kya}. In the context of our calorimetric neutrino energy reconstruction, biases can occur either through a non-linear \Ne response as a function of neutrino energy, or through an overall shift of the energy scale parameter $\langle \Ne/\Enu\rangle$. The left panel of Fig.~\ref{fig:EnergyReconstructionNeutronVariations} suggests that inefficient detection and mis-modelling of neutrons may contribute significantly to neutrino energy reconsruction biases, and therefore requires dedicated efforts in neutron simulation, reconstruction and calibration campaigns.

The right panel of Fig.~\ref{fig:EnergyReconstructionNeutronVariations} shows the neutrino energy resolution as a function of neutrino energy for fiducial $\nu_e$ CC events for the same two cases, namely the detection of all neutron-induced visible energy, or the detection of only its prompt ($<$10~$\mu$s) component. The latter assumption yields a neutrino energy resolution that is more than a factor of two worse than the former assumption, across all relevant neutrino energies. It is therefore clear that neutron detection plays a major role for accurate calorimetric neutrino energy resolution in DUNE.

In the following, we assume the more efficient neutron detection scenario for our studies. This is because the DUNE readout times per event will be at least 2.25~ms \cite{Acciarri:2016ooe}, far larger than the 10~$\mu$s detection time window considered in the other scenario. While closer to reality, our assumption may be somewhat too optimistic, considering that neutron thermalization times are also $\mathcal{O}$(1~ms) and that the hit finding algorithm for neutron-induced isolated hits may not be perfectly efficient.


\subsection{Comparison with DUNE CDR assumptions}
\label{subsec:energyreconstruction_cdr}

The energy resolution functions for $\nu_e$, $\bar{\nu}_e$, $\nu_{\mu}$, $\bar{\nu}_{\mu}$ CC interactions obtained in Sec.~\ref{subsec:energyreconstruction_larsoft} can be compared with the assumptions made in the DUNE CDR \cite{Acciarri:2015uup} and in other physics studies \cite{Adams:2013qkq,Ankowski:2015jya}. These studies rely on a different strategy to evaluate the neutrino energy resolution compared to ours. Full simulations are employed only to evaluate nuclear effects in neutrino interactions (item 1 in Tab.~\ref{tab:energyreconstruction}), but not for the LAr-TPC detector response. Rather, a parametrized response for each single particle emerging from the target nucleus is typically applied to account for all other effects (items 2--7 in Tab.~\ref{tab:energyreconstruction}). These parametrizations introduce single-particle energy and angular smearing, single-particle energy thresholds for detection and single-particle energy reconstruction biases. The parametrizations are informed from GEANT4 \cite{Agostinelli:2002hh} particle trajectories in liquid argon, by detector response simulation studies in other LAr-TPCs, by results reported by the ICARUS Collaboration, and by the expected DUNE far detector geometry, see \cite{Acciarri:2015uup,Adams:2013qkq} for details. The single-particle detector responses are then combined to estimate the reconstruction performance for event-level kinematic quantities such as the incoming neutrino energy. 

In LBNE studies \cite{Adams:2013qkq}, the $\nu_{e}$ energy resolution value is assumed to be dominated by the hadronic shower resolution. The latter is taken to be 30\%$/\sqrt{E}$, where the hadronic energy $E$ is expressed in GeV, and hadrons are assumed to carry a 40\% fraction of the total event energy on average. As a result, LBNE physics studies assume a 15\%$/\sqrt{E_{\nu}}$ resolution for $\nu_e$ CC interactions of energy $E_{\nu}$. For $\nu_{\mu}$ CC interactions, LBNE studies further assume a 10--15\% resolution for partially contained muons, to be added to the hadronic shower resolution contribution, resulting in a 20\%$/\sqrt{E_{\nu}}$ resolution overall for $\nu_{\mu}$ CC interactions. 

The DUNE CDR \cite{Acciarri:2015uup} adopts a more sophisticated, Fast MC-based, approach compared to LBNE studies. A different resolution function is assumed for several final state particle categories: muons, charged pions, electrons or gammas, protons, neutrons, and other particles. For exiting particles, the particle energy is smeared according to deposited (and not total) energy. A bias in the energy reconstruction of neutrons is also considered, with only 60\% of the neutron deposited energy assumed to be reconstructed on average. Furthermore, low-energy ($<$1~GeV/c) neutrons are considered to have a 10\% chance to remain undetected. 

In the work by Ankowski {\it et al.} \cite{Ankowski:2015jya}, all neutrons are assumed to escape detection. Two scenarios are considered. In the so-called {\it perfect reconstruction}, all particles (except neutrons) exiting the target nucleus are observed, and their measured energies are equal to the true ones. In the {\it realistic reconstruction}, the measured energies and angles are smeared to account for detector resolution effects, and single-particle detection efficiencies and energy thresholds are taken into account. We consider both scenarios in our comparison.

\begin{figure}[tbp]
\centering
\includegraphics[width=.49\textwidth]{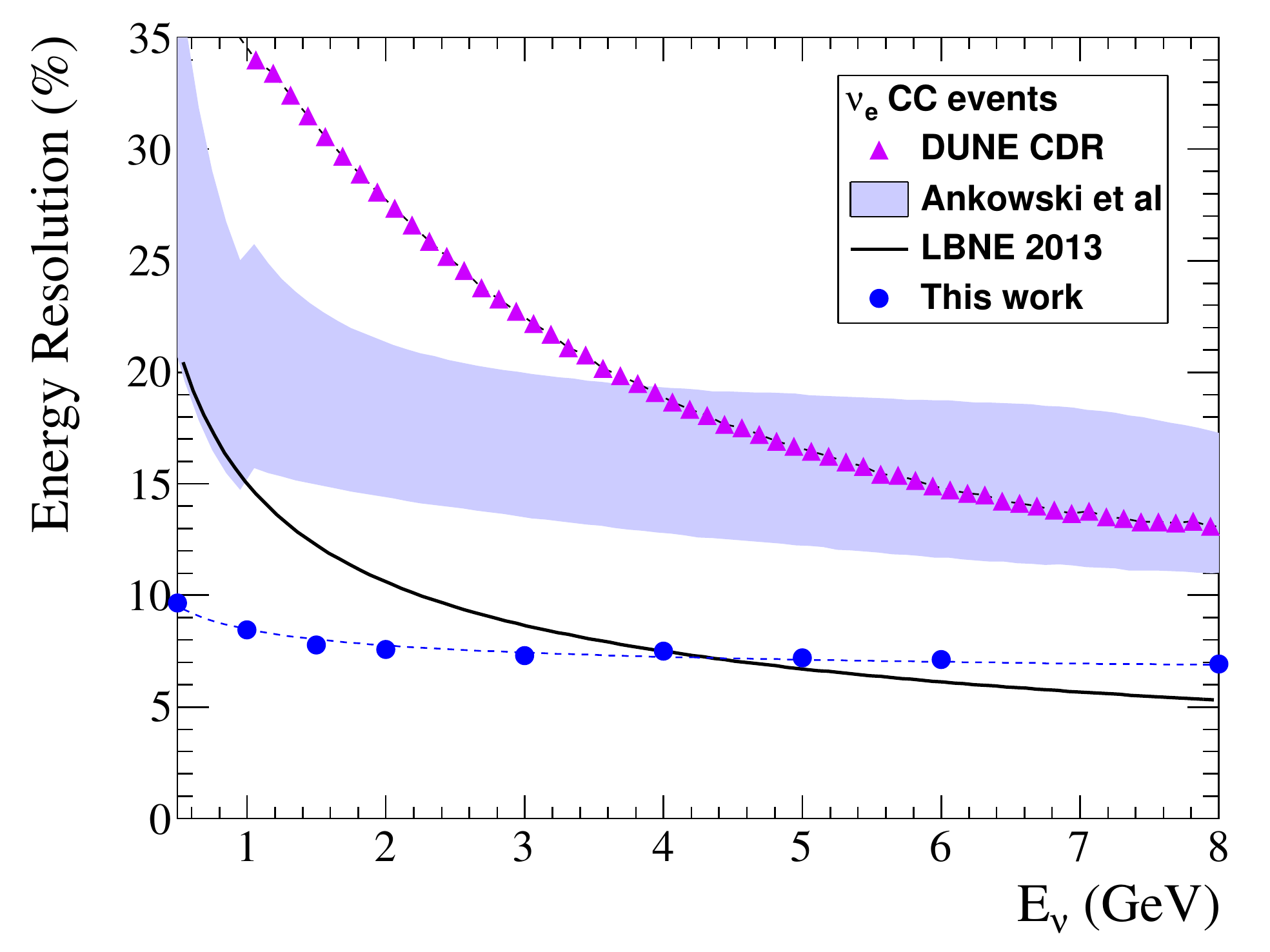} \hfill
\includegraphics[width=.49\textwidth]{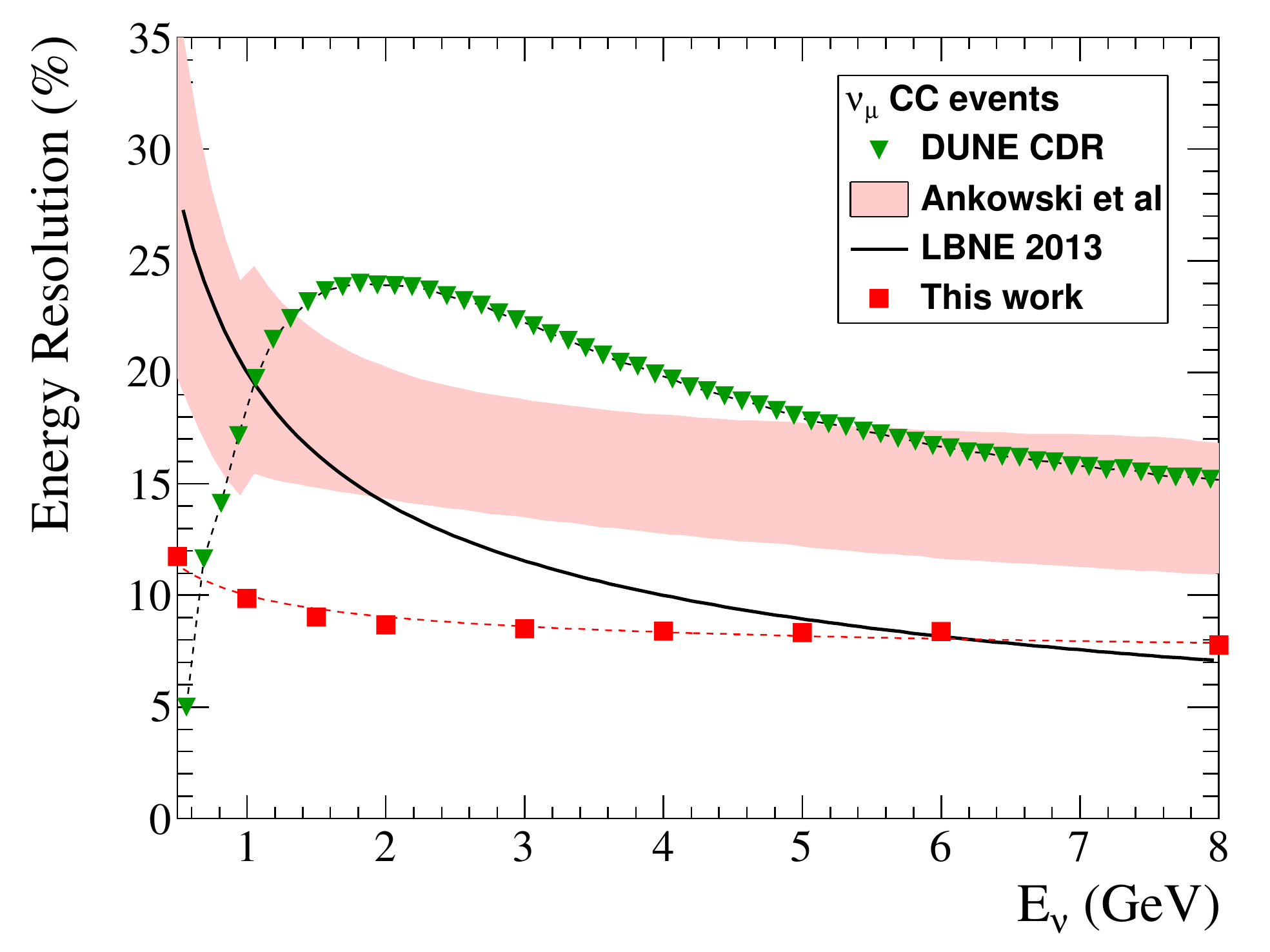}
\caption{\label{fig:EnergyResolutionComparisons}Neutrino energy resolution as a function of neutrino energy, for $\nu_{e}$ (left panel) and $\nu_{\mu}$ (right) CC interactions. Expectations from our LArSoft-based studies are shown for fiducial neutrino interactions with fully contained primary leptons. For comparison, the solid curves are the resolution functions assumed in LBNE sensitivity studies \cite{Adams:2013qkq}, the triangular markers are the DUNE CDR \cite{Acciarri:2015uup} assumptions, and the filled bands are the results obtained by Ankowski {\it et al.} \cite{Ankowski:2015jya}. For the Ankowski {\it et al.} results, the lower limit corresponds to the perfect reconstruction case, and the upper limit to the realistic reconstruction scenario.}
\end{figure}

The LBNE, DUNE CDR and Ankowski {\it et al.} assumptions for $\nu_e$ and $\nu_{\mu}$ CC interactions are shown in Fig.~\ref{fig:EnergyResolutionComparisons}, together with our LArSoft-based results. Significant differences in energy resolution assumptions exist between LBNE and DUNE CDR studies, the latter being significantly more conservative. For example, for 1~GeV $\nu_e$ CC interactions, LBNE studies assume a 15\% resolution, while the DUNE CDR about 35\% (see left panel of Fig.~\ref{fig:EnergyResolutionComparisons}). Also, our detailed simulations indicate a largely improved energy resolution at all energies compared to the DUNE CDR, and also a better resolution compared to LBNE at low energies. For example, for the same case of 1~GeV $\nu_e$ CC interactions, we expect about a 8.5\% resolution, that is about a factor of 4 better than DUNE CDR assumptions. Large improvements exist also in the $\nu_{\mu}$ CC case. Finally, our neutrino energy resolutions are also significantly better than the calorimetric resolutions obtained by Ankowski {\it et al.}. 

In order to take full advantage of these improved energy resolution assumptions, in our oscillation analysis we choose 50~MeV wide bins in reconstructed neutrino energy. This is to be compared with the 125~MeV wide bins considered in the DUNE CDR studies.


\subsection{Neutral-current neutrino interactions}
\label{subsec:energyreconstruction_nc}

Neutral-current interactions can constitute a background for $\nu_e$ appearance and $\nu_{\mu}$ disappearance searches in DUNE. For this reason, energy reconstruction studies have been performed for NC interactions as well. In the NC case, a simple gaussian smearing of true to reconstructed neutrino energy is a poor approximation, given that the outgoing neutrino carries a large fraction of the event energy. In this case, we use a more general approach. For a given true neutrino energy, the probability to obtain a reconstructed neutrino energy value is encoded in a matrix form, by using the distribution of deposited energies expected from neutrino NC interactions occurring throughout the LAr fiducial volume. Since the dominant effect on energy resolution is given in this case by non-deposited energy carried away by neutrinos, we neglect in the NC case fluctuations associated with the ionisation yield modelling and consider only items 1--3 in Tab.~\ref{tab:energyreconstruction}.

\begin{figure}[tbp]
\centering~
\includegraphics[width=.48\textwidth]{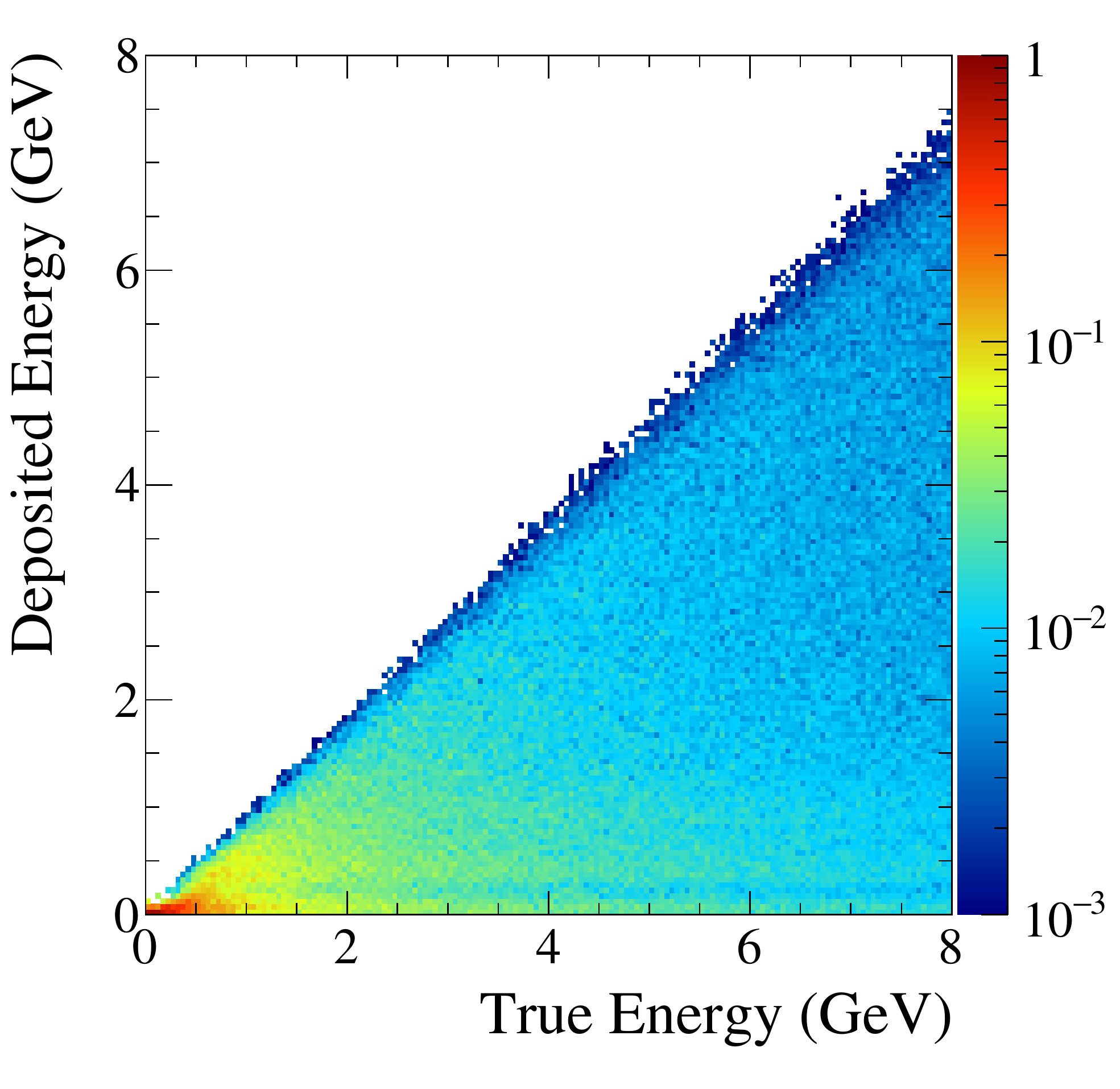} \hfill
\includegraphics[width=.48\textwidth]{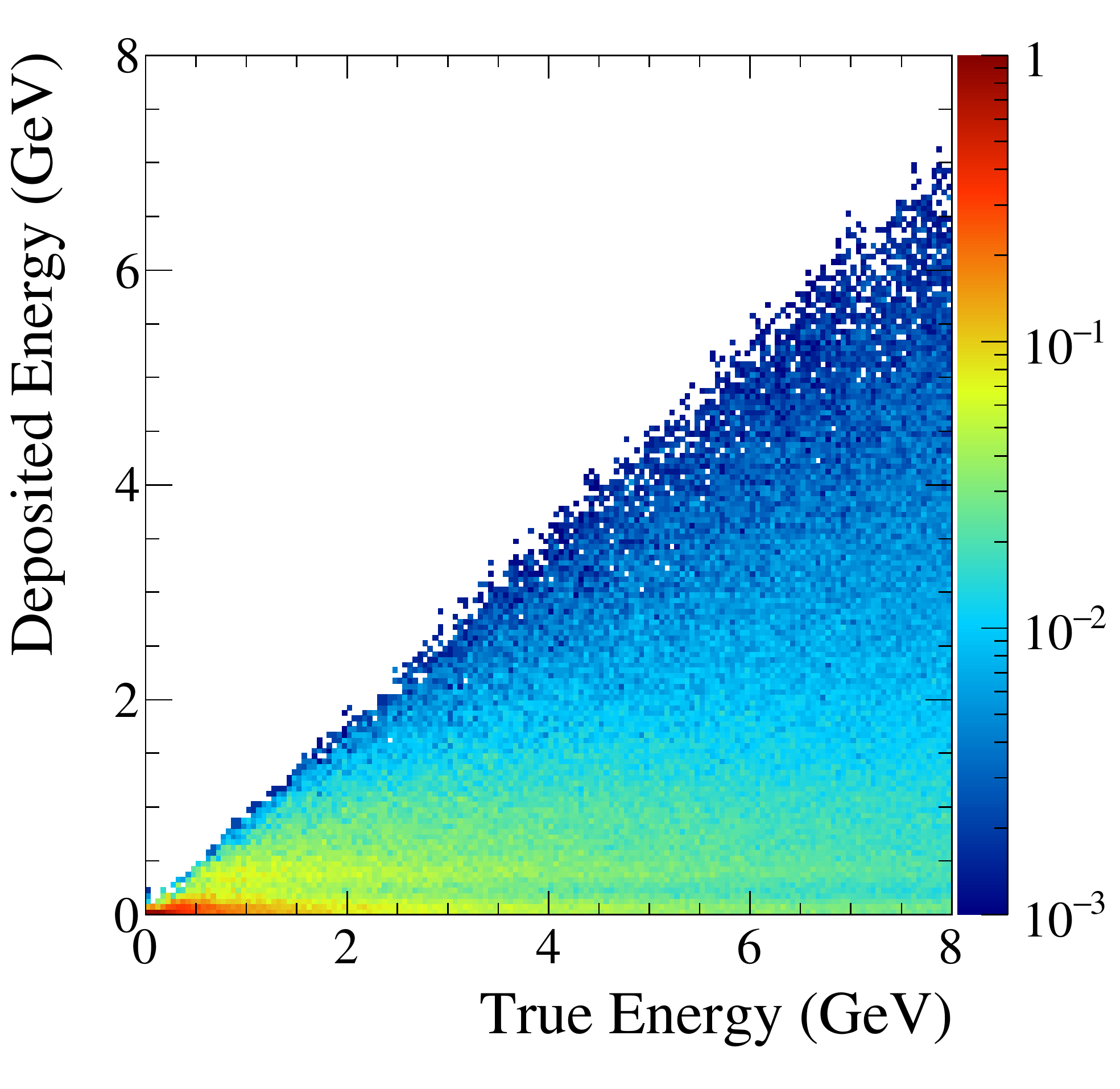} 
\caption{\label{fig:SmearingMatrixNumuNC}The expected energy smearing matrix for $\nu_{\mu}$ NC (left) and $\bar{\nu}_{\mu}$ NC (right) interactions occurring in the LAr fiducial volume. The deposited (true) neutrino energy is shown in the vertical (horizontal) axes, respectively.}
\end{figure}

 Our LArSoft-based results for neutrino and anti-neutrino NC interactions are shown separately in the left and right panels of Fig.~\ref{fig:SmearingMatrixNumuNC}. We use the smearing matrices in Fig.~\ref{fig:SmearingMatrixNumuNC} to compute the NC background event rates in our 50~MeV wide bin analysis. We have verified that these smearing matrices give comparable background rates with respect to the NC smearing assumed in the DUNE CDR \cite{Acciarri:2015uup}, when the same 125~MeV wide bins in reconstructed neutrino energy are used for both.

%% file: setup.tex
\section{Experimental setup}
\label{sec:setup}

We use the neutrino energy reconstruction results of Sec.~\ref{sec:energyreconstruction} as input to our long-baseline neutrino oscillation physics studies. We use the GLoBES software (\cite{Huber:2004ka,Huber:2007ji}, version 3.2.16) to describe the experiment and to extract oscillation sensitivities. All experimental assumptions used in our GLoBES description are given in Sec.~\ref{subsec:setup_globes}. The resulting event rates and energy spectra are presented in Sec.~\ref{subsec:setup_spectra}.


\subsection{GLoBES description}
\label{subsec:setup_globes}

Our framework to describe the DUNE experimental setup starts from the same assumptions used in the DUNE CDR \cite{Acciarri:2015uup}, which is also GLoBES-based. The experiment simulation configurations used in the DUNE CDR are described in detail in, and are publicly available at~\cite{dunecdrconfigs}. The experimental setup is thus only summarized in the following. We depart from the DUNE official description only for aspects related to neutrino energy reconstruction (the aspect we wish to address in this work) and event selection efficiencies.

The neutrino flux assumptions are the same as in the DUNE CDR \cite{Acciarri:2015uup}. They were produced with the G4LBNF Geant4-based \cite{Agostinelli:2002hh} simulation of the LBNF beamline. We use the ``optimized'' beam design in \cite{dunecdrconfigs} for our studies. The optimized design provides a significantly higher neutrino flux near the second oscillation maximum compared to the ``reference'' design. In order to account for Earth matter effects on neutrino propagation, we use a constant matter density profile from \cite{Dziewonski:1981xy,stacey}.

We assume a LAr detector fiducial mass of 40~kt, as in the DUNE CDR \cite{Acciarri:2015uup}. This fiducial target mass corresponds to four detector modules with 10~kt fiducial mass each. Overall, we assume an exposure of 150~MW$\cdot$kt$\cdot$yr in neutrino (forward horn current, FHC) running mode, and the same exposure for anti-neutrino (reverse horn current, RHC) running mode. Given the beam power (1.07~MW) and detector fiducial mass (40~kt) assumptions, these numbers correspond to about 3.5 years of FHC beam, plus 3.5 years of RHC beam \cite{dunecdrconfigs}.

Neutrino-argon interaction cross-sections in tabulated form are also taken from \cite{dunecdrconfigs}. The cross-section files are generated with the GENIE event generator code (\cite{Andreopoulos:2009rq}, version 2.8.4). Both CC and NC interactions are simulated by GENIE. The event rates assume the neutrino fluxes described above, convoluted with the oscillation probabilities and the neutrino cross-sections. As a result, realistic estimates of $\nu_e$, $\bar{\nu}_e$, $\nu_{\mu}$, $\bar{\nu}_{\mu}$, $\nu_{\tau}$, $\bar{\nu}_{\tau}$ interactions are obtained for each running mode. 

Our assumptions for energy reconstruction capabilities of the DUNE far detector are discussed in Sec.~\ref{sec:energyreconstruction}. Two scenarios are considered. First, the ``CDR scenario'' where the energy smearing matrices for neutrino CC and NC interactions of the DUNE CDR are used \cite{dunecdrconfigs}. The CC smearing matrices assume the energy-dependent resolutions shown in Figs.~\ref{fig:EnergyResolutionComparisons} for $\nu_e$ and $\nu_{\mu}$ CC events. In the CDR scenario, 125~MeV wide bins in reconstructed neutrino energy are used. We have checked that essentially no gains are to be obtained by using finer energy bins in this case. Second, we consider a scenario with improved energy reconstruction capabilities, as motivated by this work. In this case, we parametrize the energy response of CC interactions with the gaussian resolution functions of Eqs.~(\ref{eq:nueccres}--\ref{eq:numubarccres}). Furthermore, we use 50~MeV wide bins in reconstructed neutrino energy, to take full profit of the improved energy resolution assumptions. In the improved energy reconstruction scenario, we use the NC smearing matrices obtained from our LArSoft studies, see Fig.~\ref{fig:SmearingMatrixNumuNC}.

The strategy to extract oscillation parameters in DUNE is to perform a simultaneous fit to four far detector energy spectra: (1) $\nu_e$ or $\bar{\nu}_e$ CC candidate events in FHC running mode, (2) $\nu_e$ or $\bar{\nu}_e$ CC candidate events in RHC running mode, (3) $\nu_{\mu}$ or $\bar{\nu}_{\mu}$ CC candidate events in FHC running mode, (4) $\nu_{\mu}$ or $\bar{\nu}_{\mu}$ CC candidate events in RHC running mode. In our GLoBES description, we consider the same 32 oscillation channels as in the DUNE CDR, accounting for all signal and background components expected after selection cuts. We define 10 oscillation channels for the $\nu_{\mu}\to\nu_e$ appearance mode and 6 oscillation channels for the $\nu_{\mu}\to\nu_{\mu}$ disappearance mode, for both FHC and RHC running. The 32 oscillation channels are described in Tab.~2 of \cite{dunecdrconfigs}.

Efficiencies to detect and reconstruct signal and background events are also taken from the DUNE CDR \cite{Acciarri:2015uup}. They are extracted using the DUNE Fast MC \cite{Adams:2013qkq}. Efficiencies as a function of reconstructed neutrino energy for all 32 oscillation channels are provided in \cite{dunecdrconfigs}. Compared to DUNE CDR assumptions, and only for our improved energy reconstruction scenario, we further select fiducial neutrino interactions with fully contained primary leptons (see Sec.~\ref{subsec:energyreconstruction_larsoft}). We account for this by multiplying the DUNE CDR CC efficiencies by the efficiencies for primary lepton containment shown in Fig.~\ref{fig:ContainedLeptonEfficiency}.

Finally, and in addition to statistical uncertainties, we also consider the same signal and background normalization systematic uncertainties as in the DUNE CDR. We refer the reader to \cite{Acciarri:2015uup, dunecdrconfigs} for details. The DUNE CDR treatment of systematic uncertainties is customized on a per-channel basis, and requires the use of GLoBES v3.2.16 or later.


\subsection{Event rates and energy spectra}
\label{subsec:setup_spectra}

We conclude our description of the DUNE experimental setup by discussing the expected event rates in the DUNE far detector, for various event samples. We use GLoBES to extract the event rates. Each of the four far detector event samples is obtained under our two different detector performance assumptions, namely the CDR scenario and the improved energy reconstruction scenario. We summarize here the differences between the two descriptions:
\begin{enumerate}
\item Energy reconstruction for neutrino CC and NC interactions: our LArSoft-based studies (Secs.~\ref{subsec:energyreconstruction_larsoft} and \ref{subsec:energyreconstruction_nc}) versus DUNE CDR assumptions;
\item Oscillation analysis binning: 50~MeV versus 125~MeV wide bins in reconstructed neutrino energy;
\item Selection efficiency for $\nu_{\mu}$ and $\bar{\nu}_{\mu}$ CC interactions: additional primary muon full containment requirement, versus no additional requirement compared to CDR efficiencies. 
\end{enumerate}

Unless otherwise noted, the following values for the oscillation parameters from Nu-Fit \cite{Gonzalez-Garcia:2014bfa} are assumed: $\Delta m^2_{21}=(7.50\pm 0.19)\cdot 10^{-5}$~eV$^2$, $\Delta m^2_{31}=(2.457\pm 0.047)\cdot 10^{-3}$~eV$^2$, $\theta_{12}= 33.48^\circ \pm 0.78^\circ$, $\theta_{13}=8.50^\circ$ and $\theta_{23}=42.3^\circ$. The final results, unless stated, are marginalized over all other oscillation parameters within their allowed priors. In the case of $\theta_{13}$ and $\theta_{23}$ the gaussian prior representing our current constraints on these parameters are rather included for $\sin^2 2 \theta$, which is a more accurate description of the present situation and, in the case of $\theta_{23}$, allows to properly account for the octant degeneracy. Again following Nu-Fit \cite{Gonzalez-Garcia:2014bfa} we take 0.005 and 0.02 as the present errors on $\sin^2 2 \theta_{13}$ and $\sin^2 2 \theta_{23}$ respectively. Finally, a 2\% uncertainty in the PREM density profile assumed has also been considered. In general, results will either be presented as a function of the less known parameters $\delta_{CP}$, $\theta_{23}$ and the mass hierarchy or as bands whose width is determined by varying these parameters in their currently allowed range. 

\begin{table}[tbp]
\caption{\label{tab:rates}Event rates for the four DUNE far detector samples, categorized in terms of oscillation channels. Bold-face numbers indicate signal events. The $\nu_e$ CC and $\nu_{\mu}$ CC samples refer to events in FHC running mode passing the appearance and disappearance selection cuts, respectively. An exposure of 150~MW$\cdot$kt$\cdot$yr, NH and $\delta_{CP}=0$ are assumed. The $\bar{\nu}_e$ CC and $\bar{\nu}_{\mu}$ labels indicate the corresponding RHC running mode samples, with the same assumptions. Events in the 0.5--8~GeV and  0.5--20~GeV energy ranges are considered for the $\stackrel{(-)}{\nu}_e$ and $\stackrel{(-)}{\nu}_{\mu}$ samples, respectively. Numbers outside parentheses are for the CDR scenario, numbers inside parentheses are for the improved energy reconstruction scenario.}
\centering
\begin{tabular}{|l|rl|rl|rl|rl|} \hline
                         & \multicolumn{8}{|c|}{Sample} \\ \hline
Channel & \multicolumn{2}{|c|}{$\nu_e$ CC} & \multicolumn{2}{|c|}{$\bar{\nu}_e$ CC} & \multicolumn{2}{|c|}{$\nu_{\mu}$ CC} & \multicolumn{2}{|c|}{$\bar{\nu}_{\mu}$ CC} \\ \hline
$\stackrel{(-)}{\nu}_{\mu}\to\stackrel{(-)}{\nu}_e$ CC     & {\bf 955.0} & (983.8) & {\bf 215.2}  & (218.9) & 0.0          & (0.0)          & 0.0  & (0.0) \\   
$\stackrel{(-)}{\nu}_e\to\stackrel{(-)}{\nu}_e$ CC         & 203.6      & (197.1) & 104.9        & (97.6)  & 0.0           &  (0.0)        & 0.0 &  (0.0) \\  
$\stackrel{(-)}{\nu}_{\mu}\to\stackrel{(-)}{\nu}_{\mu}$ CC  & 2.8        & (1.8)   & 1.6          & (0.7)   & {\bf 8439.8}  & {\bf (6546.9)} & {\bf 4164.0} & {\bf (2978.8)} \\
$\stackrel{(-)}{\nu}_{\mu}\to\stackrel{(-)}{\nu}_{\tau}$ CC & 18.7       & (10.7)  & 10.8         & (4.8)   & 29.2          & (31.4)         & 18.2  & (17.0) \\  
$\stackrel{(-)}{\nu}$ NC                                  & 17.5       & (20.0)  & 8.9          & (8.3)   & 76.2          & (47.3)         & 40.7 & (21.4) \\ \hline 
\end{tabular}
\end{table}

Table~\ref{tab:rates} shows the event rates integrated over neutrino energy, for the various samples and for the two experimental descriptions considered. For an exposure of 150~MW$\cdot$kt$\cdot$yr in FHC running mode, approximately $10^3$ $\nu_e+\bar{\nu}_e$ CC candidates are expected at the far detector, together with about $10^4$ $\nu_{\mu}+\bar{\nu}_{\mu}$ CC candidates. For the same exposure, the event statistics is significantly lower in RHC running mode. For the CDR scenario, the signal and background rates in Tab.~\ref{tab:rates} match those given in \cite{Acciarri:2015uup}. For the improved energy reconstruction scenario, lower $\nu_{\mu}+\bar{\nu}_{\mu}$ CC rates are expected because of the additional full muon containment requirement. As Tab.~\ref{tab:rates} shows, this efficiency loss is relatively mild, at the 20--30\% level, for events in the entire 0.5--20~GeV energy range, and even lower in the low-energy portion (0.5--6~GeV) that is most relevant for neutrino oscillation measurements.

\begin{figure}[tbp]
\centering
\includegraphics[width=.49\textwidth]{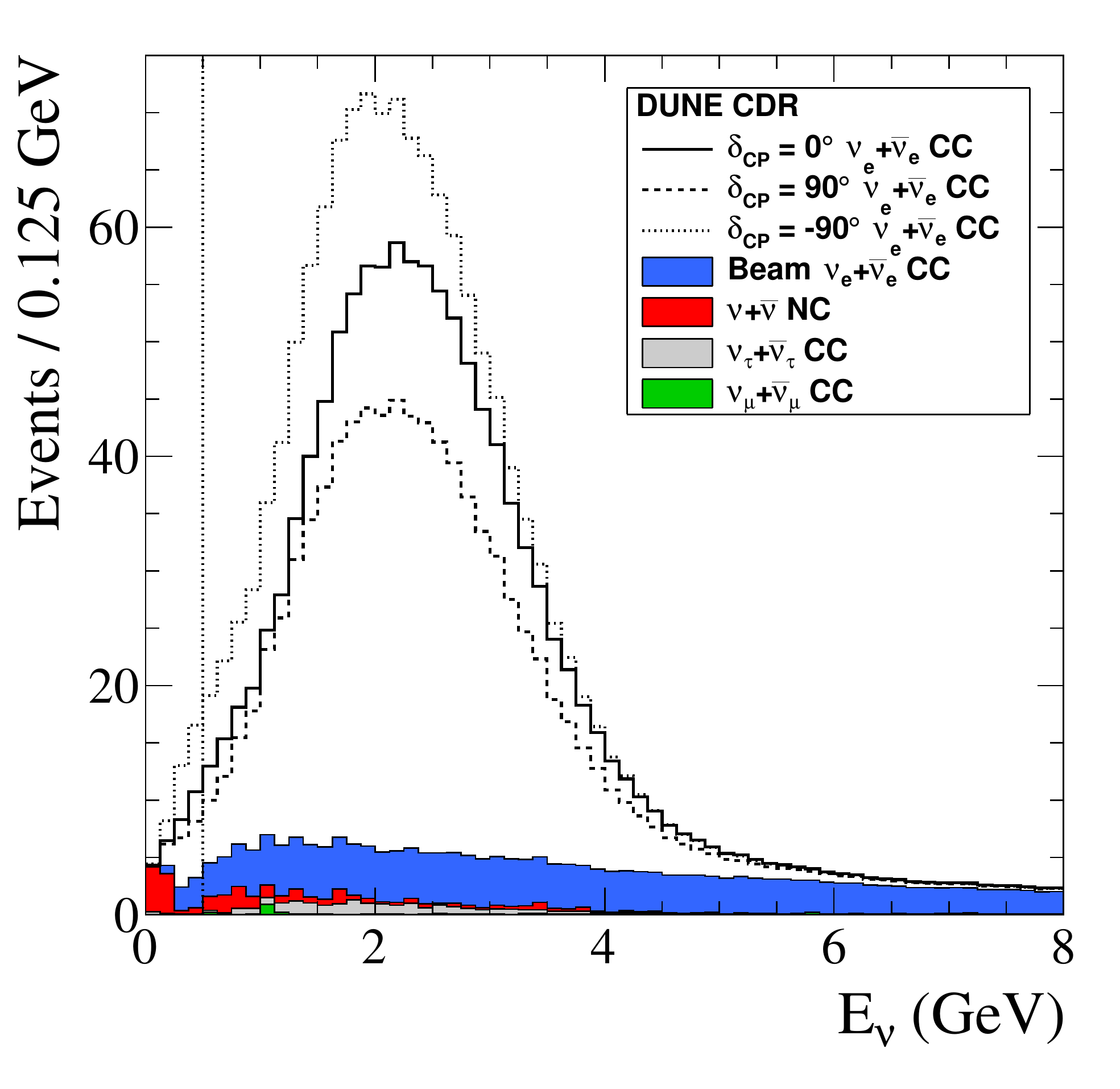} \hfill
\includegraphics[width=.49\textwidth]{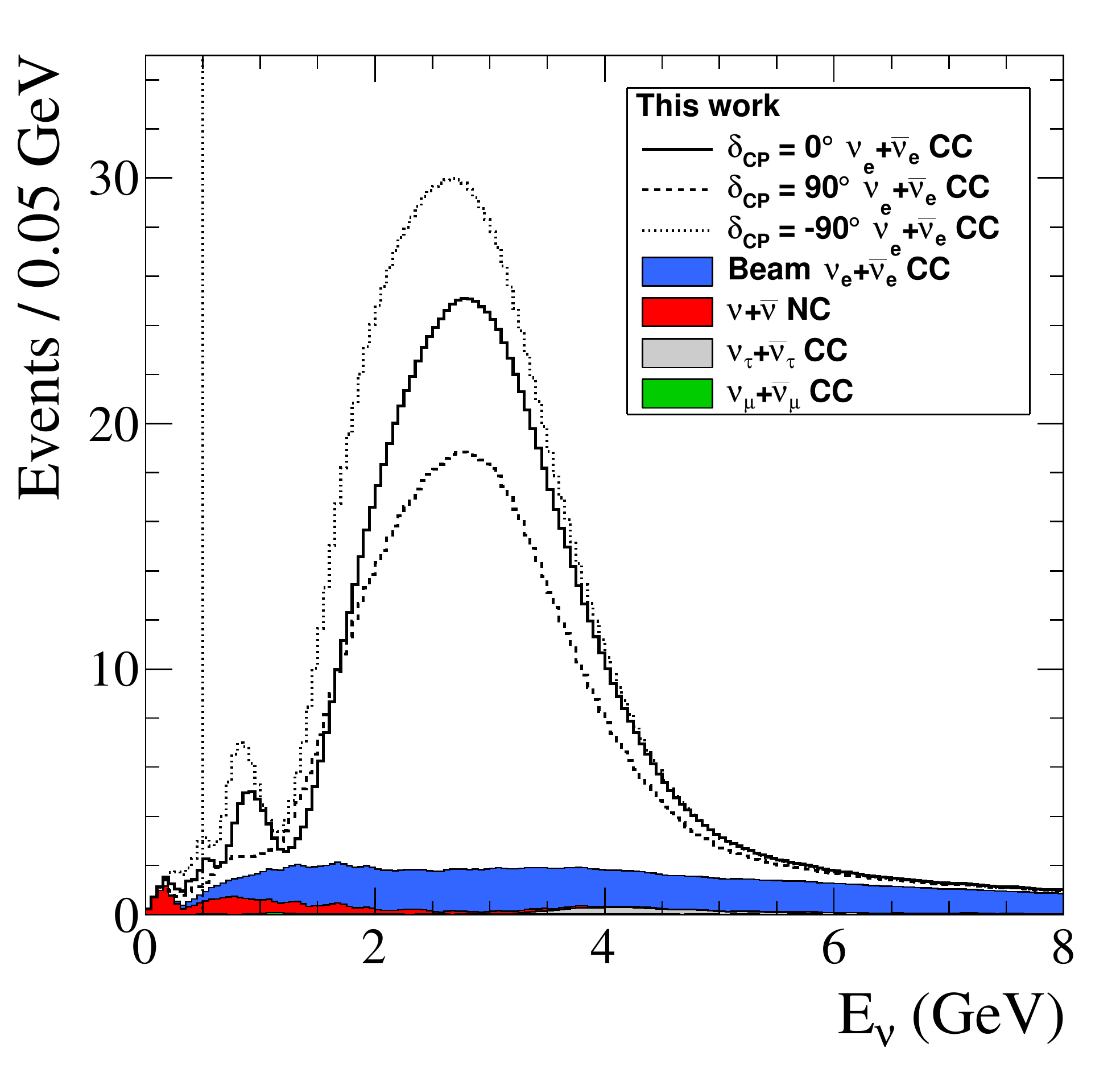}
\includegraphics[width=.49\textwidth]{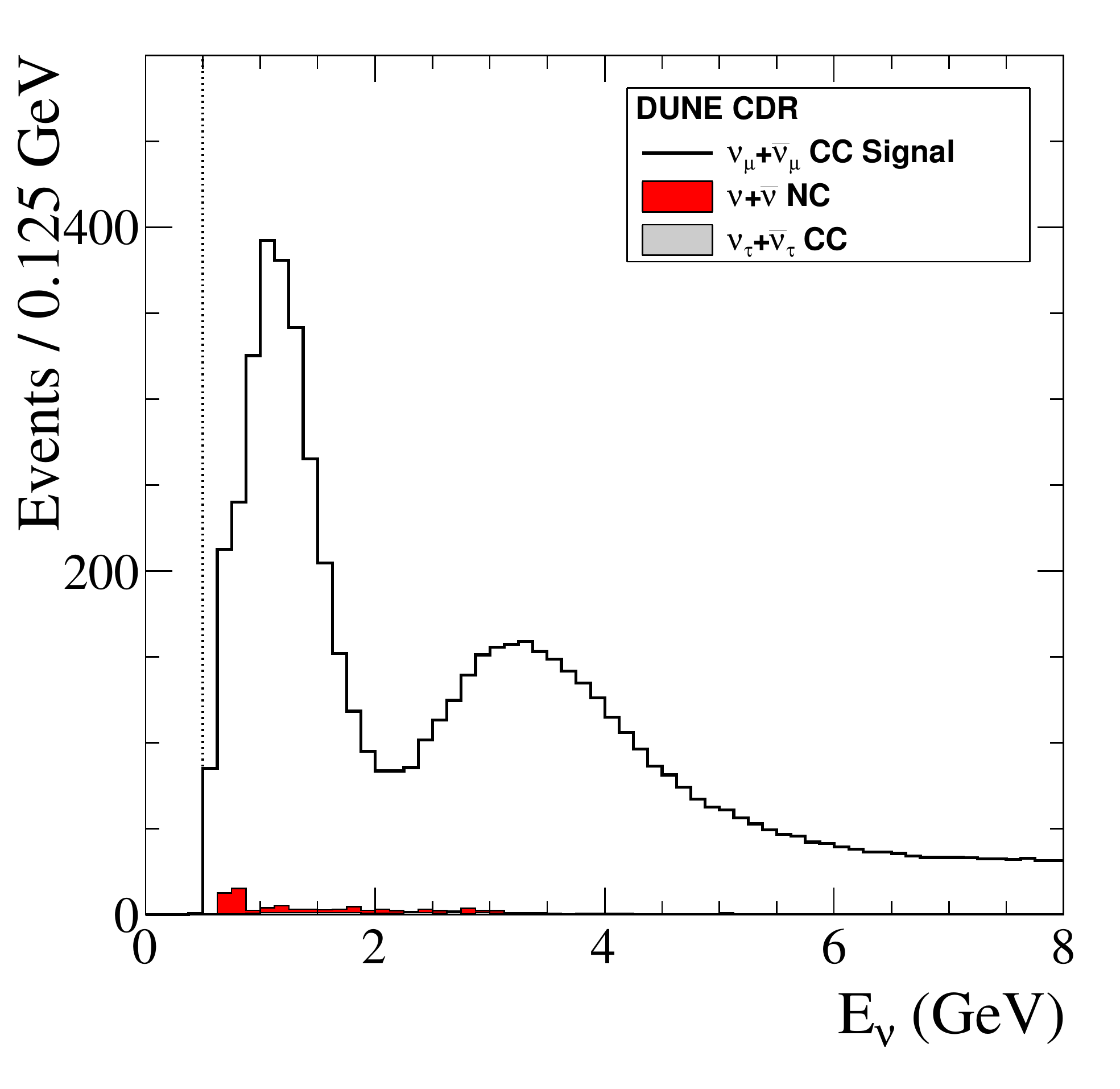} \hfill
\includegraphics[width=.49\textwidth]{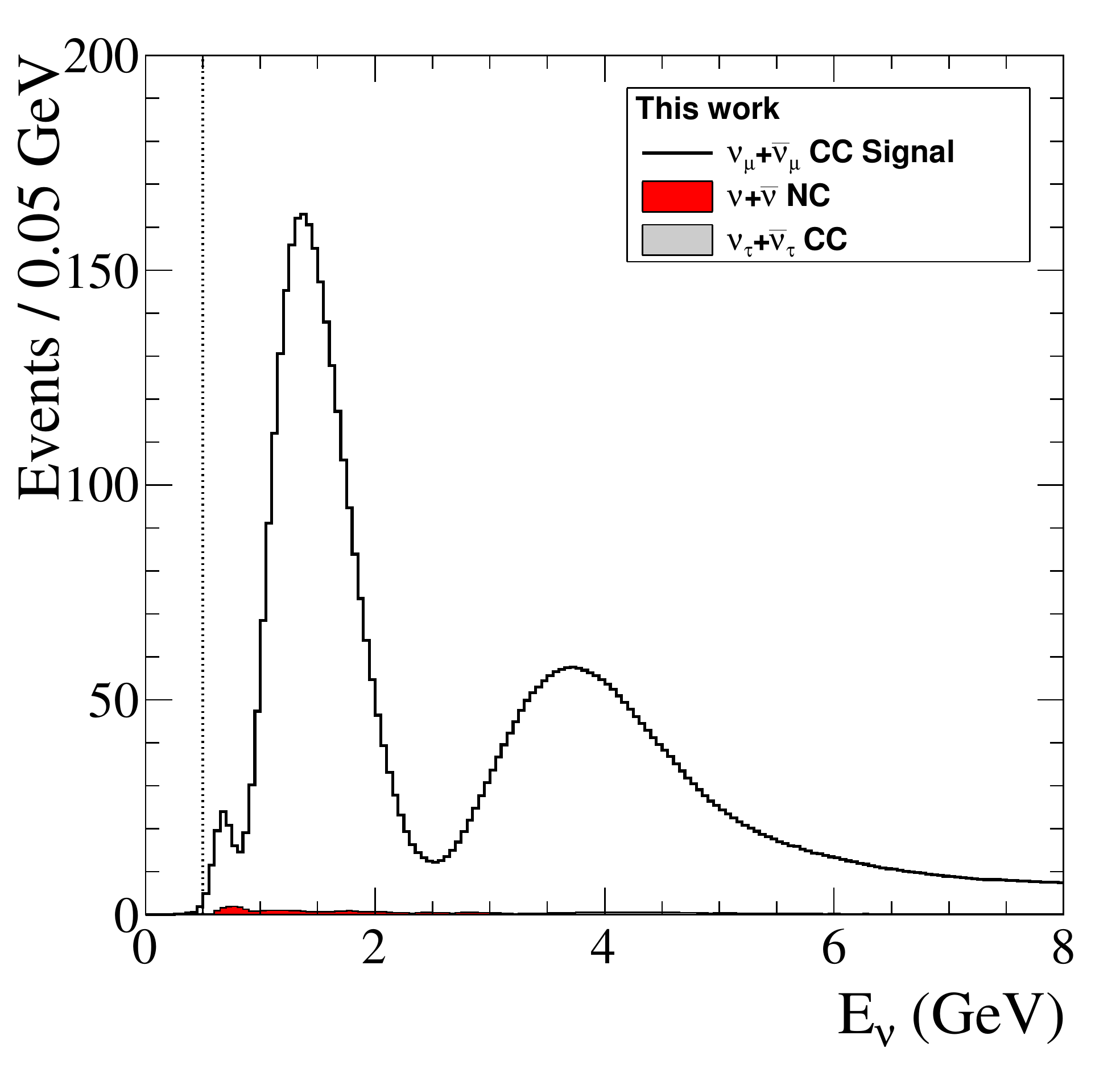}
\caption{\label{fig:Rates}Expected number of $\nu_e+\bar{\nu}_e$ CC (upper panels) and $\nu_{\mu}+\bar{\nu}_{\mu}$ CC (lower panels) candidate events in FHC running mode and as a function of reconstructed neutrino energy. Background events are shown with filled histograms, signal events with lines. Three different $\nu_e$ signal scenarios are shown in the upper panels. The neutrino energy reconstruction capabilities of the DUNE CDR \cite{Acciarri:2015uup} are assumed in the left panels, while the reconstruction performance expected from this work are shown on the right ones. An exposure of 150~MW$\cdot$kt$\cdot$yr and NH are assumed.}
\end{figure}

The expected $\nu_e+\bar{\nu}_e$ CC candidate events in FHC running mode and as a function of reconstructed neutrino energy are shown in the upper panels of Fig.~\ref{fig:Rates}, for the two scenarios described above, DUNE CDR (left) and our own assumptions (right). In the case of improved energy resolution and finer binning, a much improved determination of the second oscillation maximum at 0.8~GeV appears possible. A similar result is obtained by examining the expected $\nu_e+\bar{\nu}_e$ CC candidate events in RHC running mode. 

The $\nu_{\mu}+\bar{\nu}_{\mu}$ CC candidate event predictions in FHC mode and as a function of energy are shown in the lower panels of Fig.~\ref{fig:Rates}. An improved determination of the first oscillation dip near 2.5~GeV, and even a hint for a second oscillation dip around 0.8~GeV, appear possible with the improved reconstruction and finer analysis binning case shown in the right panel. As noted above, the efficiency loss for full muon containment has a small effect on the event rates across the relevant energy range. Again, similar conclusions are obtained for RHC mode spectra.

In the rest of this paper, we address how these improvements in spectral information translate in improvements in long-baseline oscillation sensitivities.


%% file: results.tex
\section{Results}
\label{sec:results}

The search for the remaining unknown neutrino oscillation parameters, in particular for the CP-violating phase $\delta_{CP}$, requires the study of genuinely three-flavor oscillation effects through the so-called ``golden channel'' $\nu_{\mu,e} \rightarrow \nu_{e,\mu}$. 
Neglecting higher-order terms in  $\alpha \equiv
|\Delta m_{21}^2|/|\Delta m_{31}^2| \sim 0.03$ and in $\sin \theta_{13} \sim 0.15$, the oscillation probability of $\nu_{\mu}
\rightarrow \nu_{e}$ for DUNE can be cast as~\cite{Cervera:2000kp,Freund:2001pn}:

\begin{eqnarray}
P(\nu_\mu \rightarrow \nu_e)  &\cong&  \sin^2 \theta_{23} \sin^2 2 \theta_{13} \frac{\sin^2[\Delta(A-1)] }{(A-1)^2} \nonumber\\
&+& \alpha J \sin \delta_{CP}  \sin \Delta \frac{\sin(A \Delta) \sin [(1-A)\Delta]}{A(1-A)}  \nonumber\\
&+& \alpha J \cos \delta_{CP}  \cos \Delta \frac{\sin(A \Delta) \sin [(1-A)\Delta]}{A(1-A)}  \nonumber\\
&+& \alpha^2 \cos^2 \theta_{23} \sin^2 2 \theta_{12} \frac{\sin^2 (A \Delta)}{A^2}, \label{eqn:osc_prob}
\end{eqnarray}
where 
 $J= \cos \theta_{13} \sin 2 \theta_{13} \sin 2 \theta_{12} \sin 2 \theta_{23}$, $ \Delta = \Delta m^2_{31} L/4E, A=2\sqrt{3} G_F n_e E/\Delta m^2_{31}$,  $G_F$ is the weak coupling constant and $n_e$ is the number density of electrons in the propagation medium.
An asymmetry in neutrino versus antineutrino oscillations is induced both by the presence of a CP-odd term ($\propto \sin \delta_{CP}$) and by the matter effect $A$, which changes sign going from $\nu_{\mu} \rightarrow \nu_{e}$ to the $\bar{\nu}_{\mu} \rightarrow \bar{\nu}_{e}$ channel.
The asymmetry induced by the matter effect depends on the sign of  $\Delta m^2_{31}$ and increases with the neutrino energy $(E)$ and the baseline $(L)$. The degeneracy between the CP-violation and matter effect induced asymmetries can thus be resolved  
with long baselines and high energies, like the ones adopted by DUNE. This experiment enhances so much the matter effect that the associated asymmetry cannot be mimicked by any possible value of $\delta_{CP}$. In this case the CP-violating asymmetry would represent a far more sub-leading - and therefore challenging to discover - effect. 

On the other hand, these strong matter effects not only lead to a great enhancement of the  (anti)neutrino channel for NH (IH), but also to the suppression of the oscillation probability for the other channel. This implies that the search for leptonic CP violation cannot rely so strongly on its most natural physics effect: an asymmetry between the neutrino and antineutrino oscillation probabilities - since one of them is very suppressed; it must rather exploit the characteristic energy dependence of the CP-violating term in Eq.~(\ref{eqn:osc_prob}). 

Furthermore, the fact that $\theta_{13}$ turned out to be relatively large, saturating previous upper bounds, implies that the expansion in Eq.~(\ref{eqn:osc_prob}) is not symmetric since the terms suppressed by $\sin \theta_{13} \sim 0.15$ dominate over those suppressed by $\alpha \sim 0.03$. In particular, the first term of the equation tends to dominate over the second and third, which are those containing the dependence on $\delta_{CP}$. A possible way to alleviate this is to observe the oscillation probability beyond the first oscillation peak, so that the slower, $\Delta m^2_{21}$-driven oscillation has developed further and the CP-violating interference with the $\sin \theta_{13}$-modulated term represents a more significant contribution to the final oscillation probability~\cite{Diwan:2003bp,Barger:2006vy,Barger:2007yw}. Thus, one of the widely-advertised benefits of a wide-band beam - such as the one envisioned for the DUNE experiment - is that it would allow to cover not only the first oscillation peak, but also beyond it and, in particular, the second oscillation maximum. However, in Ref.~\cite{Huber:2010dx} it was shown that for a wide-band beam, the significantly lower statistics present at the second oscillation peak due to the smaller flux (typically at the tail) and cross section (at lower energies), combined with the large background from NC interactions migrating from higher energies, rendered the information coming from the second peak nearly irrelevant (see also \cite{Ghosh:2014rna})\footnote{An interesting alternative for experiments with high statistics is to center the beam energy at the second oscillation peak, which can significantly increase the sensitivity to $\delta_{CP}$~\cite{Coloma:2011pg,Baussan:2013zcy,Agarwalla:2014tpa}.}. Thus, an improved energy reconstruction can be beneficial to the search for leptonic CP violation in two ways. On the one hand it helps with the measurement of $\delta_{CP}$ through the characteristic energy dependence of the second and third terms in Eq.~(\ref{eqn:osc_prob}). On the other hand, it can enhance the relevance of the measurements at the second oscillation peak as shown in Fig.~\ref{fig:Rates}, particularly when combined with the optimized beam design recently adopted, which peaks at lower energies.

In the rest of this section we will present sensitivity studies for: (1) the MH, (2) $\delta_{CP}$ and (3) $\theta_{23}$. All experimental sensitivities are estimated using the GLoBES package \cite{Huber:2004ka}, and are obtained for the DUNE experiment description and for the priors in the neutrino oscillation parameters described in Sec.~\ref{sec:setup}.

\subsection{Sensitivity to mass hierarchy}

The neutrino mass hierarchy, that is the sign of $\Delta m^2_{31}$, controls the sign of the matter potential $A$ in Eq.~(\ref{eqn:osc_prob}) and it can thus lead to an enhancement (suppression) of the neutrino oscillation probability for normal (inverted) hierarchy, with an opposite effect for antineutrino oscillations. This effect is already present (and quadratic) in the first term of Eq.~(\ref{eqn:osc_prob}) which, given that $\sin \theta_{13} > \alpha$, turns out to dominate the oscillation probability. Thus, for the long-baseline and high neutrino energies that characterize the DUNE setup, large matter effects, and hence very good sensitivity to the neutrino mass hierarchy, are expected.  

We quantify the MH discovery potential through the $\chi^2$ difference between the two hierarchies:
\begin{eqnarray}
\Delta\chi^2_{\rm MH} & = & \chi^2_{\rm IH} - \chi^2_{\rm NH}, ~\rm (for~ true~ normal ~hierarchy) \nonumber \\
\Delta\chi^2_{\rm MH} & = & \chi^2_{\rm NH} - \chi^2_{\rm IH}, ~\rm (for~ true~ inverted ~hierarchy).
\end{eqnarray}
Notice that some deviations from a $\chi^2$ distribution are expected for this observable, however, these have been quantitatively evaluated to be small~\cite{Blennow:2013oma}.
 
The sensitivity of DUNE to the MH determination depends on the actual values of $\delta_{CP}$ and $\theta_{23}$, apart from the true value of the MH itself.
On the one hand, the role of $\delta_{CP}$ is similar to that of the MH since it induces an asymmetry in the neutrino vs antineutrino oscillation probabilities. Indeed, NH and $\delta_{CP} \sim -\pi/2$ will both enhance (suppress) the (anti)neutrino oscillations, while an IH and $\delta_{CP} \sim \pi/2$ would have the opposite effect. Thus, any of these two combinations would lead to the best sensitivities to both the MH and the CP-violating phase, while cases with NH and $\delta_{CP} \sim \pi/2$ or IH and $\delta_{CP} \sim -\pi/2$ tend to compensate each other leading to an altogether weaker effect and lower sensitivities. Regarding $\theta_{23}$, since the sensitivity to the MH mainly stems from the first term in Eq.~(\ref{eqn:osc_prob}), proportional to $\sin^2 \theta_{23}$,  larger values of $\theta_{23}$ are preferred to enhance the effect and increase the sensitivity.  For these reasons, we perform a scan over all possible values of $\delta^{\rm true}_{CP}$ and show the results for a range of true values of $\theta_{23}$ allowed by current global fits \cite{Forero:2014bxa,Capozzi:2013csa,Gonzalez-Garcia:2014bfa}. 

\begin{figure}[!htbp]
\centering
\includegraphics[width=0.99\textwidth]{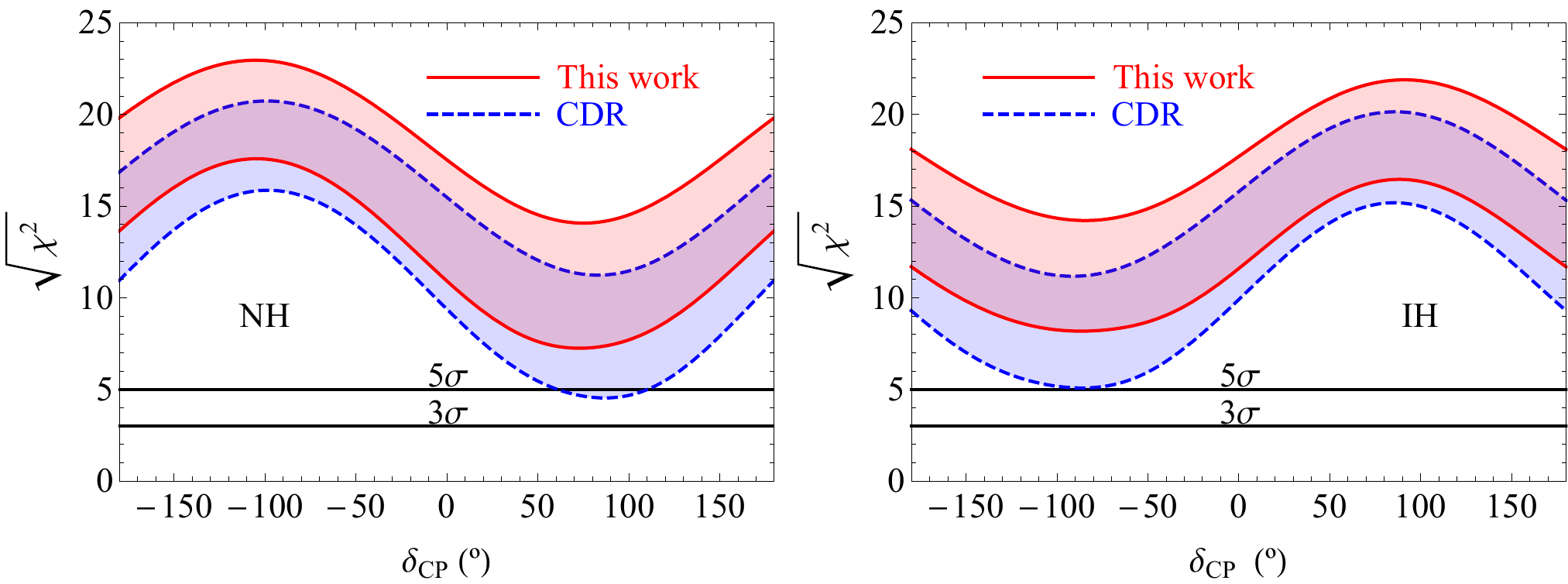}
\caption{\label{fig:hiersens} Mass hierarchy discovery potential for DUNE at a fiducial exposure of 1.07~MW$\times$40~kton$\times$7~yr. The plot on the left is for NH, the one on the right is for IH. Blue, dashed curves refer to the significance with which the MH can be determined, as a function of $\delta_{CP}$, for the DUNE CDR setup.  Red, solid curves show the expected sensitivity for the improved energy reconstruction discussed in this work. The width of the band corresponds to letting the assumed true value of $\theta_{23}$ vary from $\theta_{23} = 38^{\circ}$ (lower sensitivity) to $\theta_{23} = 53^{\circ}$ (higher sensitivity) as allowed by current global fits.}
\end{figure}

Figure~\ref{fig:hiersens} shows the significance (under the assumption of a $\chi^2$ distribution) with which the MH could be determined as a function of the value of $\delta^{\rm true}_{CP}$, for an exposure of 3.5 + 3.5 years (in FHC and RHC modes, respectively), with a 40~kt detector and a 1.07~MW beam. With this exposure, the MH may be determined with a minimum significance of $\sqrt{\Delta\chi^2_{\rm MH}} = 5$ for any $\delta_{CP}$ value for the DUNE CDR detector performance scenario (blue dashed curves) except for NH close to $\delta_{CP} \sim \pi/2$ and for small $\theta_{23}$ where the significance is slightly over $4 \sigma$. This sensitivity is enhanced by at least $2 \sigma$ for the improved energy reconstruction scenario motivated by this paper (red solid curves). Even the least favorable combination of $(\theta_{23},\delta_{CP},\mbox{MH})$ parameters would allow to disentangle the MH  with a significance of $\sim 7\sigma$ in the improved reconstruction scenario.

\subsection{Sensitivity to CP violation and precision in $\delta_{CP}$}

DUNE aims at the observation of leptonic CP violation (CPV). Within the standard framework of three neutrino flavours, such a signal will be observable if the value of the CP violating phase is sufficiently different from 0 or $\pi$, cases in which there is no CP violation. Furthermore, the sensitivity to $\delta_{CP}$ stems solely from the second and third terms in Eq.~(\ref{eqn:osc_prob}). Therefore, smaller values of $\theta_{23}$, which suppress the dominant first term in the equation, tend to increase the sensitivity to leptonic CP violation. We infer the CP violation discovery potential of DUNE at a given value $\delta_{CP}^{\rm true}$  by minimising the $\Delta\chi^2$: 
	
\begin{eqnarray}
\Delta\chi^2_{\rm CPV} & = & \min\left[\Delta\chi^2_{CP}(\delta_{CP}^{\rm test}=0),\Delta\chi^2_{CP}(\delta_{CP}^{\rm test}=\pi)\right]\mbox{, where} \\
\Delta\chi^2_{\rm CP} & = & \chi^2_{\delta_{CP}^{\rm test}} - \chi^2_{\delta_{CP}^{\rm true}}. \\ \nonumber
\end{eqnarray}
Even though the cyclic nature of $\delta_{CP}$ has been shown to induce important deviations from a $\chi^2$ distribution, these are mainly relevant for present facilities~\cite{Schwetz:2006md,Blennow:2014sja,Elevant:2015ska} and their impact for the DUNE setup should be mild.

\begin{figure}[!htbp]
\centering
\includegraphics[width=0.99\textwidth]{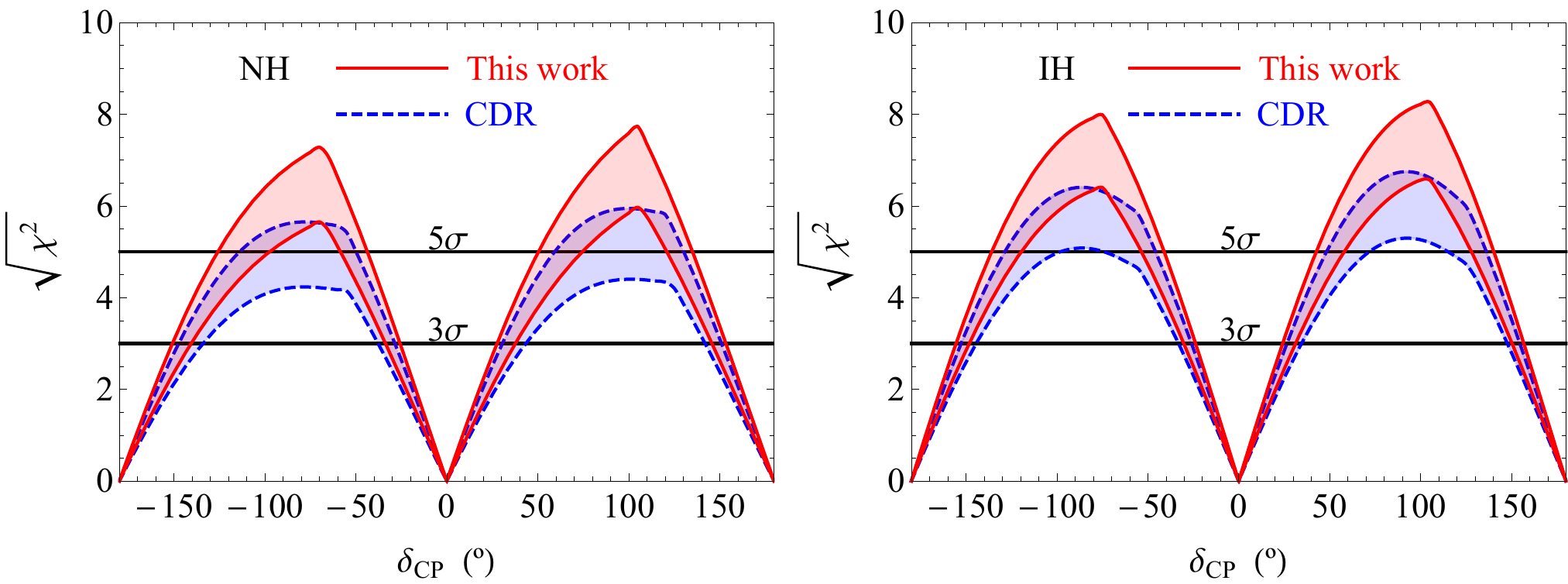} \\
\includegraphics[width=0.99\textwidth]{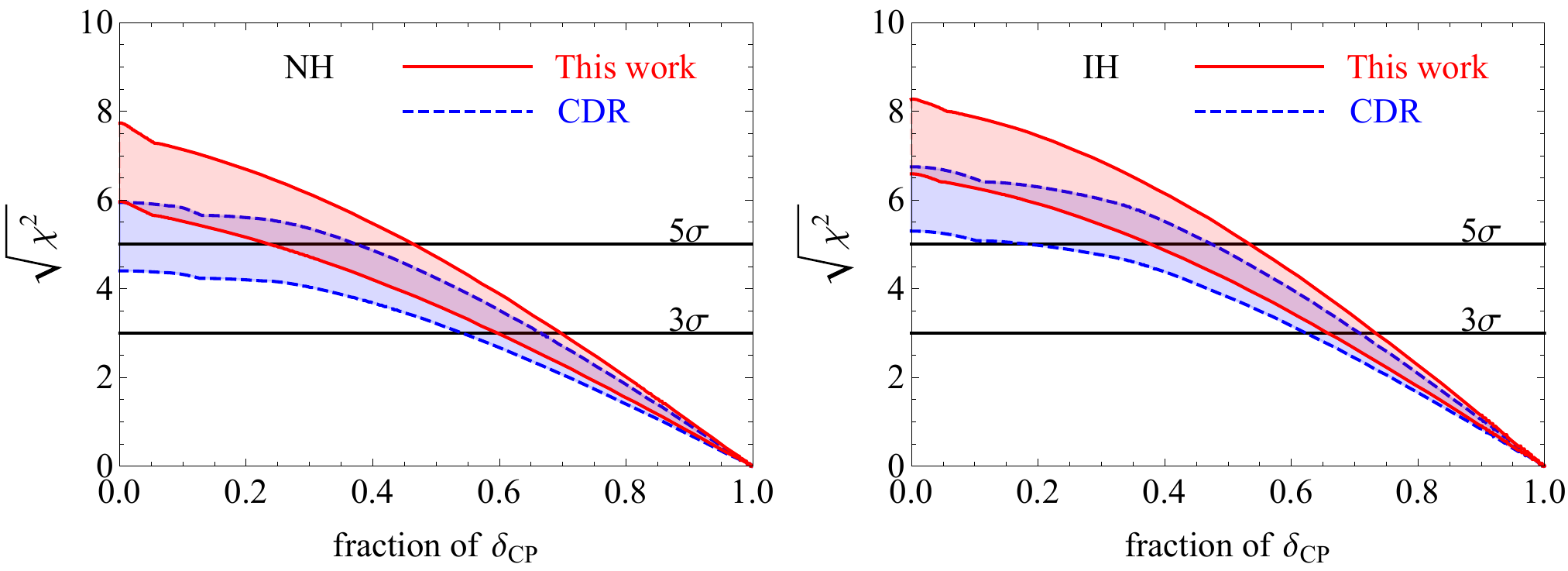}
\caption{\label{fig:deltaCP} CP violation discovery potential ($\delta_{CP} \neq 0, \pi$) at DUNE, assuming a fiducial exposure of 1.07~MW$\times$40~kton$\times$7~yr, as a function of $\delta_{CP}$. The plot on the left is for NH, the one on the right is for IH. Blue, dashed curves refer to the DUNE CDR setup.  Red, solid curves show the expected discovery potential for the improved energy reconstruction discussed in this work. The width of the band corresponds to letting the assumed true value of $\theta_{23}$ vary from $\theta_{23} = 53^{\circ}$ (lower sensitivity) to $\theta_{23} = 38^{\circ}$ (higher sensitivity) as allowed by current global fits. The lower panels convey the same information, but as a function of the fraction of values of $\delta_{CP}$ for which a given significance could be achieved.}
\end{figure}

Sensitivities as a function of $\delta_{CP}^{\rm true}$ are shown in Fig.~\ref{fig:deltaCP} for a range of true values of $\theta_{23}$ allowed by current global fits, for both NH and IH. Solid black lines corresponding to $3\sigma$ and $5\sigma$ significance, which indicate 99.73\%  and 99.99\% probability respectively, of determining CP-violation (under a $\chi^2$ distribution assumption) are also shown for comparison. We also show in the lower panels the same information, but as a function of the fraction of values of $\delta_{CP}$ that would allow a discovery of CPV at the corresponding significance.

For the DUNE CDR scenario, a $3 \sigma$ evidence for CPV could be achieved for $\sim 60 \%$ ($\sim 65 \%$) of the possible values of $\delta_{CP}$ for NH (IH), depending on the actual value of $\theta_{23}$. These values would increase by $\sim 5 \%$ with the improved energy reconstruction discussed here. The improvement is however more dramatic for the $5 \sigma$ discovery mark. In this case, under the CDR assumptions, a maximum of  $37 \%$ ($47 \%$) of the possible values of $\delta_{CP}$ for NH (IH) would be covered for favorable values of $\theta_{23}$, while no values of $\delta_{CP}$ would allow for such a discovery for the least favorable $\theta_{23}$ values. With improved energy reconstruction, these numbers are improved to $47 \%$ ($53 \%$), and a minimum $24 \%$ coverage is obtained even for the least favorable $\theta_{23}$ values. 

\begin{figure}[!htbp]
\centering
\includegraphics[width=0.99\textwidth]{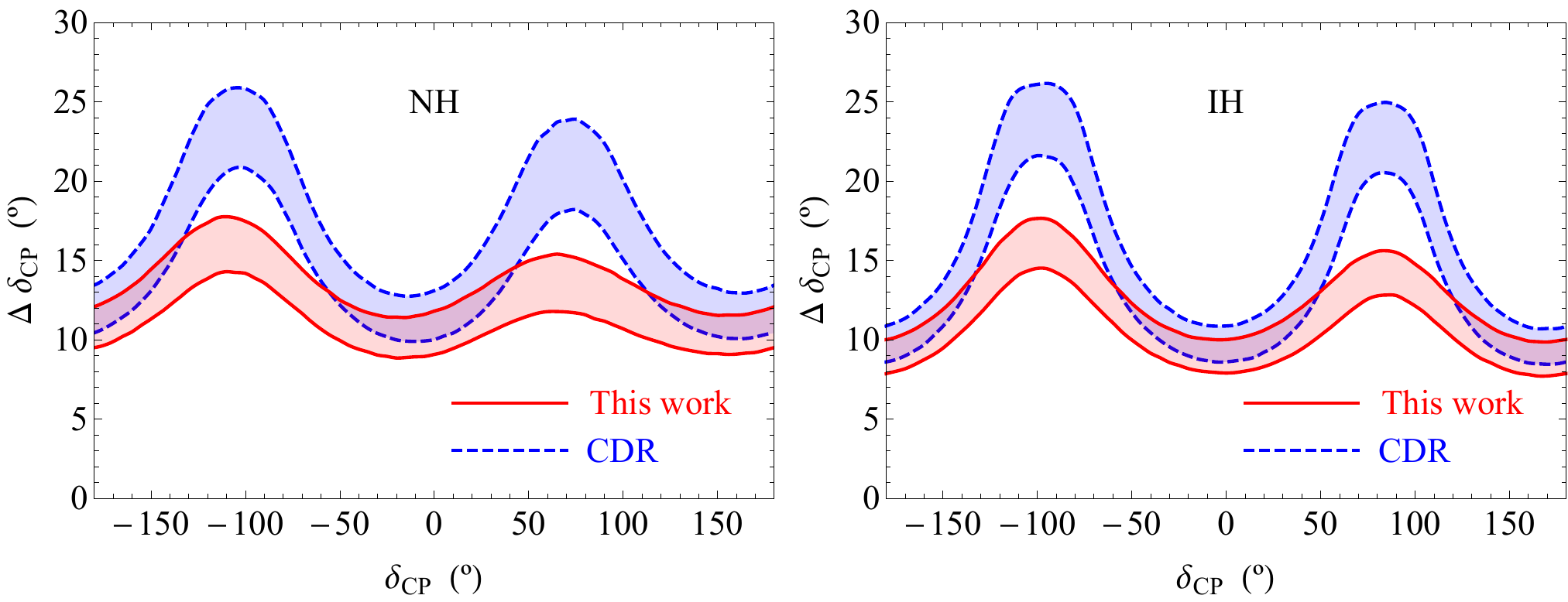}
\caption{\label{fig:precCP} Half-size of the allowed $1 \sigma$ region to which $\delta_{CP}$ would be constrained at DUNE, assuming a fiducial exposure of 1.07~MW$\times$40~kton$\times$7~yr, as a function of $\delta_{CP}$. The plot on the left is for NH, the one on the right is for IH. Blue, dashed curves refer to the DUNE CDR setup.  Red, solid curves show the expected discovery potential for the improved energy reconstruction discussed in this work. The width of the band corresponds to letting the assumed true value of $\theta_{23}$ vary from $\theta_{23} = 53^{\circ}$ (smaller error) to $\theta_{23} = 38^{\circ}$ (larger error) as allowed by current global fits.}
\end{figure}

In Fig.~\ref{fig:precCP} we show the $\delta_{CP}$ uncertainties, defined as the half-size of the $1 \sigma$ region to which $\delta_{CP}$ would be constrained at DUNE. For the CDR setup (blue dashed curves) this error would range between $9^\circ$ and $26^\circ$. The smallest uncertainties correspond to small values of $\theta_{23}$ (minimizing the competing CP-conserving term in Eq.~(\ref{eqn:osc_prob})) and to nearly CP-conserving $\delta_{CP}$ values, although slightly shifted to the left due to matter effects (see Ref.~\cite{Coloma:2012wq} for a detailed explanation of this effect). The improved energy reconstruction greatly reduces the $\delta_{CP}$ uncertainties around the maximally CP-violating values of $\delta_{CP}$, translating into a $\delta_{CP}$ precision ranging between $8^\circ$ and $18^\circ$. This very pronounced improvement for CP-violating values of $\delta_{CP}$ is also apparent in the substantial gain in the $5 \sigma$ coverage of the CP-violation discovery potential shown in Fig~\ref{fig:deltaCP}. We have verified that these gains stem significantly from the improved observation of the second oscillation maximum thanks to the better energy resolution. Indeed, when restricting the analysis to the higher energy bins so that the second maximum in not analyzed, the improvement between the CDR scenario and the one with better energy reconstruction is notably milder. As noted above, we have also verified that the finer (50~MeV compared to 125~MeV) energy binning has no impact on the sensitivities if combined with the CDR energy reconstruction. In other words, the effect is entirely caused by the improved energy reconstruction.    

\subsection{Sensitivity to $\theta_{23}$ octant degeneracy and to deviations from maximal mixing}

Current global fit analyses~\cite{Forero:2014bxa,Capozzi:2013csa,Gonzalez-Garcia:2014bfa} find good agreement with experimental data for values of $\theta_{23}$ slightly above and below $\pi/4$. This stems from the fact that, at present, the measurements of this parameter are dominated by the $\nu_\mu$ disappearance channel, with a leading dependence on $\sin 2 \theta_{23}$, which induces this degeneracy. Furthermore, maximal mixing is only mildly disfavored in these fits and remains a perfectly viable possibility. It is thus interesting to explore the sensitivity of DUNE to rule out maximal mixing and to determine its correct octant (if $\theta_{23}$ turns out not to be maximal). This information is particularly interesting for models aiming at explaining the observed pattern of neutrino masses and mixings through symmetry arguments, since deviations of $\theta_{23}$ from maximal mixing tend to be correlated with the value of $\theta_{13}$ or even $\delta_{CP}$.  

With DUNE, the maximal mixing and the octant hypotheses can be probed via a synergistic measurement of both  $\nu_\mu \rightarrow \nu_\mu$ and $\nu_\mu \rightarrow \nu_e$  oscillations, sensitive to  $\sin^2 2\theta_{23}$ and to $\sin^2 \theta_{23}$, respectively. We define the $\Delta\chi^2$ metric in the two cases as:

\begin{eqnarray}
\Delta\chi^2_{\rm maxmix} & = &  \chi^2(\theta_{23}=45^\circ) - \chi^2(\theta_{23}^{true}), \nonumber \\
\Delta\chi^2_{\rm octant} & = &  \chi^2(\pi/2-\theta_{23}^{true}) - \chi^2(\theta_{23}^{true})  .
\end{eqnarray}
In this analysis we do not impose the gaussian prior on $\sin^2 2\theta_{23}$.

\begin{figure}[!htbp]
\centering
\includegraphics[width=0.99\textwidth]{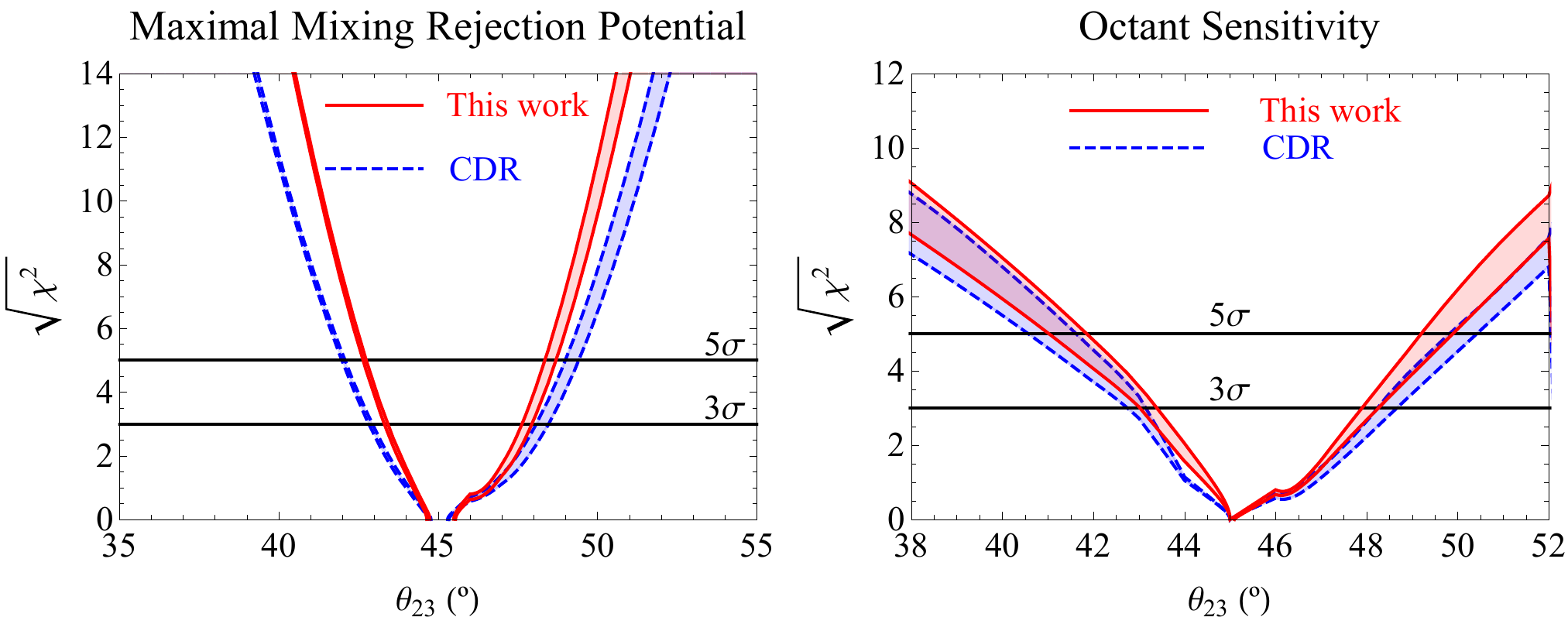}
\caption{\label{fig:maxmixNH_octant} Maximal mixing rejection potential (left) and octant discovery potential (right) of DUNE, assuming a fiducial exposure of 1.07~MW$\times$40~kton$\times$7~yr. Blue, dashed curves refer to the DUNE CDR setup.  Red, solid curves show the expected discovery potential for the improved energy reconstruction discussed in this work. The width of the band corresponds to allowing $\delta_{CP}$ to vary in its whole range as well as allowing both possible mass hierarchies.}
\end{figure}

The left panel of Fig \ref{fig:maxmixNH_octant} shows the maximal mixing rejection potential, that is, the significance (assuming a $\chi^2$ distribution) with which the maximal mixing hypothesis would be disfavored as a function of the true value of $\theta_{23}$. Values as close to $45^\circ$ as $42^\circ$ ($43^\circ$) or $50^\circ$ ($49^\circ$) for the CDR (improved energy reconstruction) setup could be distinguished from maximal mixing at $5\sigma$ significance. For comparison, the $5\sigma$ range in $\theta_{23}$ from global fits to existing neutrino oscillation data lies in the range $36^\circ$--$56^\circ$ \cite{Gonzalez-Garcia:2014bfa}. In the right panel of Fig \ref{fig:maxmixNH_octant} the significance (under the assumption of a $\chi^2$ distribution) with which the octant degeneracy could be solved is shown. For the CDR (improved energy reconstruction) setup a discovery of the octant would be guaranteed, unless $\theta_{23}$ lies in in the interval $40^\circ-51^\circ$ ($41^\circ-50^\circ$). For these two measurements the sensitivity mainly stems from the combination of the appearance and disappearance channels. Therefore, the better resolution of the second oscillation maximum and the improved energy reconstruction are less relevant, and translate in a less significant gain compared to the $\delta_{CP}$ measurement. In this case, the lower statistics in the disappearance channel for the improved energy reconstruction setup, due to the additional requirement of full muon containment, also plays a role. 

\newpage

%% file: conclusions.tex
\section{Conclusions}
\label{sec:conc}

In this work we have explored the gains in DUNE long-baseline neutrino oscillation sensitivities that can be obtained with improved neutrino energy reconstruction capabilities, compared to what is customarily assumed.

The neutrino energy reconstruction performance for neutrino charged-current interactions of all relevant flavors and energies has been evaluated via detailed simulations of a DUNE liquid argon far detector module. Our simulations account for nuclear effects in neutrino-argon interactions, energy leakage out of the detector active volume, ionization quenching, and electron-ion recombination, while we neglect drift and readout effects. Our studies indicate that an energy reconstruction performance significantly better than what was assumed in the DUNE Conceptual Design Report (CDR, \cite{Acciarri:2015uup}) is in principle possible, particularly at low neutrino energies.

We find that this improved energy reconstruction not only allows DUNE to better resolve the characteristic energy dependence of the CP-violating term. It can also allow the experiment to reconstruct much more clearly the second oscillation maximum of the $\nu_\mu \to \nu_e$ oscillation, particularly sensitive to the unknown CP-violating phase $\delta_{CP}$. In combination with DUNE's optimized neutrino beamline design, yielding a larger flux at low energies compared to the previous reference design, the improved energy reconstruction translates into a significant increase in the physics reach of the DUNE search for $\delta_{CP}$. Indeed, for an exposure of of 3.5 + 3.5 years (in forward horn current and reverse horn current modes, respectively), with a 40~kt fiducial mass detector and a 1.07~MW beam, the improved energy reconstruction translates into an increase of the fraction of $\delta_{CP}$ values allowing for a $5 \sigma$ discovery of leptonic CP violation from 0 \% (47 \%) to 24 \% (53 \%) for the least (most) favorable combination of the other neutrino oscillation parameters allowed by current data. The precision of a $\delta_{CP}$ measurement would also increase significantly, from a maximum uncertainty of $26^\circ$ with standard (DUNE CDR) detector response, to $18^\circ$ with improved energy reconstruction. The significance of the mass hierarchy determination would also increase by at least two standard deviations for any value of $\delta_{CP}$.

Conversely, the precision measurements of $\theta_{23}$ do not benefit as much from the improved energy reconstruction, relying more in the comparison of appearance and disappearance channels than in the coverage of the second oscillation maximum. Still, a modest increase of $\sim 2^\circ$ in the range of $\theta_{23}$ values that would allow for either a $5\sigma$ determination of the octant or for a rejection of maximal mixing, would be achieved. 

We expect that the benefits to the DUNE oscillation sensitivities from improved energy reconstruction could be somewhat reduced by the inclusion of energy shape systematic uncertainties. Energy shape-only systematic uncertainties have been neglected both in the DUNE CDR and in our own studies. Future analyses, adopting a more comprehensive systematic uncertainty treatment, will therefore prove to be interesting. In any case, we conclude that the observed gain in DUNE physics reach is strong enough to justify further exploration of the energy reconstruction capabilities ultimately achievable with large liquid argon neutrino detectors, and particularly with DUNE.

%% file: main.bbl
\begin{thebibliography}{64}
\expandafter\ifx\csname natexlab\endcsname\relax\def\natexlab#1{#1}\fi
\expandafter\ifx\csname bibnamefont\endcsname\relax
  \def\bibnamefont#1{#1}\fi
\expandafter\ifx\csname bibfnamefont\endcsname\relax
  \def\bibfnamefont#1{#1}\fi
\expandafter\ifx\csname citenamefont\endcsname\relax
  \def\citenamefont#1{#1}\fi
\expandafter\ifx\csname url\endcsname\relax
  \def\url#1{\texttt{#1}}\fi
\expandafter\ifx\csname urlprefix\endcsname\relax\def\urlprefix{URL }\fi
\providecommand{\bibinfo}[2]{#2}
\providecommand{\eprint}[2][]{\url{#2}}

\bibitem[{\citenamefont{Pontecorvo}(1957)}]{Pontecorvo:1957cp}
\bibinfo{author}{\bibfnamefont{B.}~\bibnamefont{Pontecorvo}},
  \bibinfo{journal}{Sov.Phys.JETP} \textbf{\bibinfo{volume}{6}},
  \bibinfo{pages}{429} (\bibinfo{year}{1957}).

\bibitem[{\citenamefont{Pontecorvo}(1958)}]{Pontecorvo:1957qd}
\bibinfo{author}{\bibfnamefont{B.}~\bibnamefont{Pontecorvo}},
  \bibinfo{journal}{Sov.Phys.JETP} \textbf{\bibinfo{volume}{7}},
  \bibinfo{pages}{172} (\bibinfo{year}{1958}).

\bibitem[{\citenamefont{Maki et~al.}(1960)\citenamefont{Maki, Nakagawa, Ohnuki,
  and Sakata}}]{Maki:1960ut}
\bibinfo{author}{\bibfnamefont{Z.}~\bibnamefont{Maki}},
  \bibinfo{author}{\bibfnamefont{M.}~\bibnamefont{Nakagawa}},
  \bibinfo{author}{\bibfnamefont{Y.}~\bibnamefont{Ohnuki}}, \bibnamefont{and}
  \bibinfo{author}{\bibfnamefont{S.}~\bibnamefont{Sakata}},
  \bibinfo{journal}{Prog.Theor.Phys.} \textbf{\bibinfo{volume}{23}},
  \bibinfo{pages}{1174} (\bibinfo{year}{1960}).

\bibitem[{\citenamefont{Maki et~al.}(1962)\citenamefont{Maki, Nakagawa, and
  Sakata}}]{Maki:1962mu}
\bibinfo{author}{\bibfnamefont{Z.}~\bibnamefont{Maki}},
  \bibinfo{author}{\bibfnamefont{M.}~\bibnamefont{Nakagawa}}, \bibnamefont{and}
  \bibinfo{author}{\bibfnamefont{S.}~\bibnamefont{Sakata}},
  \bibinfo{journal}{Prog.Theor.Phys.} \textbf{\bibinfo{volume}{28}},
  \bibinfo{pages}{870} (\bibinfo{year}{1962}).

\bibitem[{\citenamefont{Pontecorvo}(1968)}]{Pontecorvo:1967fh}
\bibinfo{author}{\bibfnamefont{B.}~\bibnamefont{Pontecorvo}},
  \bibinfo{journal}{Sov.Phys.JETP} \textbf{\bibinfo{volume}{26}},
  \bibinfo{pages}{984} (\bibinfo{year}{1968}).

\bibitem[{\citenamefont{Gonzalez-Garcia
  et~al.}(2014)\citenamefont{Gonzalez-Garcia, Maltoni, and
  Schwetz}}]{Gonzalez-Garcia:2014bfa}
\bibinfo{author}{\bibfnamefont{M.~C.} \bibnamefont{Gonzalez-Garcia}},
  \bibinfo{author}{\bibfnamefont{M.}~\bibnamefont{Maltoni}}, \bibnamefont{and}
  \bibinfo{author}{\bibfnamefont{T.}~\bibnamefont{Schwetz}},
  \bibinfo{journal}{JHEP} \textbf{\bibinfo{volume}{11}}, \bibinfo{pages}{052}
  (\bibinfo{year}{2014}), \eprint{1409.5439}.

\bibitem[{\citenamefont{An et~al.}(2012)}]{An:2012eh}
\bibinfo{author}{\bibfnamefont{F.}~\bibnamefont{An}} \bibnamefont{et~al.}
  (\bibinfo{collaboration}{DAYA-BAY Collaboration}),
  \bibinfo{journal}{Phys.Rev.Lett.} \textbf{\bibinfo{volume}{108}},
  \bibinfo{pages}{171803} (\bibinfo{year}{2012}), \eprint{1203.1669}.

\bibitem[{\citenamefont{Ahn et~al.}(2012)}]{Ahn:2012nd}
\bibinfo{author}{\bibfnamefont{J.}~\bibnamefont{Ahn}} \bibnamefont{et~al.}
  (\bibinfo{collaboration}{RENO collaboration}),
  \bibinfo{journal}{Phys.Rev.Lett.} \textbf{\bibinfo{volume}{108}},
  \bibinfo{pages}{191802} (\bibinfo{year}{2012}), \eprint{1204.0626}.

\bibitem[{\citenamefont{Abe et~al.}(2012)}]{Abe:2012tg}
\bibinfo{author}{\bibfnamefont{Y.}~\bibnamefont{Abe}} \bibnamefont{et~al.}
  (\bibinfo{collaboration}{Double Chooz Collaboration}),
  \bibinfo{journal}{Phys.Rev.} \textbf{\bibinfo{volume}{D86}},
  \bibinfo{pages}{052008} (\bibinfo{year}{2012}), \eprint{1207.6632}.

\bibitem[{\citenamefont{Adamson et~al.}(2011)}]{Adamson:2011qu}
\bibinfo{author}{\bibfnamefont{P.}~\bibnamefont{Adamson}} \bibnamefont{et~al.}
  (\bibinfo{collaboration}{MINOS Collaboration}),
  \bibinfo{journal}{Phys.Rev.Lett.} \textbf{\bibinfo{volume}{107}},
  \bibinfo{pages}{181802} (\bibinfo{year}{2011}), \eprint{1108.0015}.

\bibitem[{\citenamefont{Abe et~al.}(2011)}]{Abe:2011sj}
\bibinfo{author}{\bibfnamefont{K.}~\bibnamefont{Abe}} \bibnamefont{et~al.}
  (\bibinfo{collaboration}{T2K Collaboration}),
  \bibinfo{journal}{Phys.Rev.Lett.} \textbf{\bibinfo{volume}{107}},
  \bibinfo{pages}{041801} (\bibinfo{year}{2011}), \eprint{1106.2822}.

\bibitem[{\citenamefont{Olive et~al.}(2014)}]{Agashe:2014kda}
\bibinfo{author}{\bibfnamefont{K.~A.} \bibnamefont{Olive}} \bibnamefont{et~al.}
  (\bibinfo{collaboration}{Particle Data Group}), \bibinfo{journal}{Chin.
  Phys.} \textbf{\bibinfo{volume}{C38}}, \bibinfo{pages}{090001}
  (\bibinfo{year}{2014}).

\bibitem[{\citenamefont{Gavela et~al.}(1994{\natexlab{a}})\citenamefont{Gavela,
  Hernandez, Orloff, and Pene}}]{Gavela:1993ts}
\bibinfo{author}{\bibfnamefont{M.}~\bibnamefont{Gavela}},
  \bibinfo{author}{\bibfnamefont{P.}~\bibnamefont{Hernandez}},
  \bibinfo{author}{\bibfnamefont{J.}~\bibnamefont{Orloff}}, \bibnamefont{and}
  \bibinfo{author}{\bibfnamefont{O.}~\bibnamefont{Pene}},
  \bibinfo{journal}{Mod.Phys.Lett.} \textbf{\bibinfo{volume}{A9}},
  \bibinfo{pages}{795} (\bibinfo{year}{1994}{\natexlab{a}}),
  \eprint{hep-ph/9312215}.

\bibitem[{\citenamefont{Gavela et~al.}(1994{\natexlab{b}})\citenamefont{Gavela,
  Hernandez, Orloff, Pene, and Quimbay}}]{Gavela:1994dt}
\bibinfo{author}{\bibfnamefont{M.}~\bibnamefont{Gavela}},
  \bibinfo{author}{\bibfnamefont{P.}~\bibnamefont{Hernandez}},
  \bibinfo{author}{\bibfnamefont{J.}~\bibnamefont{Orloff}},
  \bibinfo{author}{\bibfnamefont{O.}~\bibnamefont{Pene}}, \bibnamefont{and}
  \bibinfo{author}{\bibfnamefont{C.}~\bibnamefont{Quimbay}},
  \bibinfo{journal}{Nucl.Phys.} \textbf{\bibinfo{volume}{B430}},
  \bibinfo{pages}{382} (\bibinfo{year}{1994}{\natexlab{b}}),
  \eprint{hep-ph/9406289}.

\bibitem[{\citenamefont{Abe et~al.}(2014)}]{Abe:2013hdq}
\bibinfo{author}{\bibfnamefont{K.}~\bibnamefont{Abe}} \bibnamefont{et~al.}
  (\bibinfo{collaboration}{T2K Collaboration}), \bibinfo{journal}{Phys. Rev.
  Lett.} \textbf{\bibinfo{volume}{112}}, \bibinfo{pages}{061802}
  (\bibinfo{year}{2014}), \eprint{1311.4750}.

\bibitem[{\citenamefont{Adamson et~al.}(2016{\natexlab{a}})}]{Adamson:2016tbq}
\bibinfo{author}{\bibfnamefont{P.}~\bibnamefont{Adamson}} \bibnamefont{et~al.}
  (\bibinfo{collaboration}{NOvA}), \bibinfo{journal}{Phys. Rev. Lett.}
  \textbf{\bibinfo{volume}{116}}, \bibinfo{pages}{151806}
  (\bibinfo{year}{2016}{\natexlab{a}}), \eprint{1601.05022}.

\bibitem[{\citenamefont{Capozzi et~al.}(2016)\citenamefont{Capozzi, Lisi,
  Marrone, Montanino, and Palazzo}}]{Capozzi:2016rtj}
\bibinfo{author}{\bibfnamefont{F.}~\bibnamefont{Capozzi}},
  \bibinfo{author}{\bibfnamefont{E.}~\bibnamefont{Lisi}},
  \bibinfo{author}{\bibfnamefont{A.}~\bibnamefont{Marrone}},
  \bibinfo{author}{\bibfnamefont{D.}~\bibnamefont{Montanino}},
  \bibnamefont{and} \bibinfo{author}{\bibfnamefont{A.}~\bibnamefont{Palazzo}},
  \bibinfo{journal}{Nucl. Phys.} \textbf{\bibinfo{volume}{B908}},
  \bibinfo{pages}{218} (\bibinfo{year}{2016}), \eprint{1601.07777}.

\bibitem[{\citenamefont{Gonzalez-Garcia et~al.}()\citenamefont{Gonzalez-Garcia,
  Maltoni, and Schwetz}}]{nufit2_1}
\bibinfo{author}{\bibfnamefont{M.}~\bibnamefont{Gonzalez-Garcia}},
  \bibinfo{author}{\bibfnamefont{M.}~\bibnamefont{Maltoni}}, \bibnamefont{and}
  \bibinfo{author}{\bibfnamefont{T.}~\bibnamefont{Schwetz}},
  \bibinfo{note}{{NuFIT 2.1 (2016), \url{http://www.nu-fit.org}}}.

\bibitem[{\citenamefont{Fukugita and Yanagida}(1986)}]{Fukugita:1986hr}
\bibinfo{author}{\bibfnamefont{M.}~\bibnamefont{Fukugita}} \bibnamefont{and}
  \bibinfo{author}{\bibfnamefont{T.}~\bibnamefont{Yanagida}},
  \bibinfo{journal}{Phys. Lett.} \textbf{\bibinfo{volume}{B174}},
  \bibinfo{pages}{45} (\bibinfo{year}{1986}).

\bibitem[{\citenamefont{Acciarri et~al.}(2015)}]{Acciarri:2015uup}
\bibinfo{author}{\bibfnamefont{R.}~\bibnamefont{Acciarri}} \bibnamefont{et~al.}
  (\bibinfo{collaboration}{DUNE}) (\bibinfo{year}{2015}), \eprint{1512.06148}.

\bibitem[{\citenamefont{Sorel}(2014)}]{Sorel:2014rka}
\bibinfo{author}{\bibfnamefont{M.}~\bibnamefont{Sorel}},
  \bibinfo{journal}{JINST} \textbf{\bibinfo{volume}{9}},
  \bibinfo{pages}{P10002} (\bibinfo{year}{2014}), \eprint{1405.0848}.

\bibitem[{\citenamefont{Ankowski
  et~al.}(2015{\natexlab{a}})\citenamefont{Ankowski, Benhar, Coloma, Huber,
  Jen, Mariani, Meloni, and Vagnoni}}]{Ankowski:2015jya}
\bibinfo{author}{\bibfnamefont{A.~M.} \bibnamefont{Ankowski}},
  \bibinfo{author}{\bibfnamefont{O.}~\bibnamefont{Benhar}},
  \bibinfo{author}{\bibfnamefont{P.}~\bibnamefont{Coloma}},
  \bibinfo{author}{\bibfnamefont{P.}~\bibnamefont{Huber}},
  \bibinfo{author}{\bibfnamefont{C.-M.} \bibnamefont{Jen}},
  \bibinfo{author}{\bibfnamefont{C.}~\bibnamefont{Mariani}},
  \bibinfo{author}{\bibfnamefont{D.}~\bibnamefont{Meloni}}, \bibnamefont{and}
  \bibinfo{author}{\bibfnamefont{E.}~\bibnamefont{Vagnoni}},
  \bibinfo{journal}{Phys. Rev.} \textbf{\bibinfo{volume}{D92}},
  \bibinfo{pages}{073014} (\bibinfo{year}{2015}{\natexlab{a}}),
  \eprint{1507.08560}.

\bibitem[{\citenamefont{Michael et~al.}(2008)}]{Michael:2008bc}
\bibinfo{author}{\bibfnamefont{D.~G.} \bibnamefont{Michael}}
  \bibnamefont{et~al.} (\bibinfo{collaboration}{MINOS}),
  \bibinfo{journal}{Nucl. Instrum. Meth.} \textbf{\bibinfo{volume}{A596}},
  \bibinfo{pages}{190} (\bibinfo{year}{2008}), \eprint{0805.3170}.

\bibitem[{\citenamefont{Adamson et~al.}(2016{\natexlab{b}})}]{Adamson:2016xxw}
\bibinfo{author}{\bibfnamefont{P.}~\bibnamefont{Adamson}} \bibnamefont{et~al.}
  (\bibinfo{collaboration}{NOvA}), \bibinfo{journal}{Phys. Rev.}
  \textbf{\bibinfo{volume}{D93}}, \bibinfo{pages}{051104}
  (\bibinfo{year}{2016}{\natexlab{b}}), \eprint{1601.05037}.

\bibitem[{\citenamefont{Ferrari and Sala}(2001)}]{Ferrari:2000wu}
\bibinfo{author}{\bibfnamefont{A.}~\bibnamefont{Ferrari}} \bibnamefont{and}
  \bibinfo{author}{\bibfnamefont{P.~R.} \bibnamefont{Sala}},
  \bibinfo{journal}{Frascati Phys. Ser.} \textbf{\bibinfo{volume}{21}},
  \bibinfo{pages}{31} (\bibinfo{year}{2001}), \bibinfo{note}{[,31(2000)]}.

\bibitem[{\citenamefont{Rubbia et~al.}(2011)}]{Rubbia:2011ft}
\bibinfo{author}{\bibfnamefont{C.}~\bibnamefont{Rubbia}} \bibnamefont{et~al.},
  \bibinfo{journal}{JINST} \textbf{\bibinfo{volume}{6}},
  \bibinfo{pages}{P07011} (\bibinfo{year}{2011}), \eprint{1106.0975}.

\bibitem[{\citenamefont{Stahl et~al.}(2012)}]{Stahl:2012exa}
\bibinfo{author}{\bibfnamefont{A.}~\bibnamefont{Stahl}} \bibnamefont{et~al.}
  (\bibinfo{year}{2012}).

\bibitem[{\citenamefont{Ankowski
  et~al.}(2015{\natexlab{b}})\citenamefont{Ankowski, Coloma, Huber, Mariani,
  and Vagnoni}}]{Ankowski:2015kya}
\bibinfo{author}{\bibfnamefont{A.~M.} \bibnamefont{Ankowski}},
  \bibinfo{author}{\bibfnamefont{P.}~\bibnamefont{Coloma}},
  \bibinfo{author}{\bibfnamefont{P.}~\bibnamefont{Huber}},
  \bibinfo{author}{\bibfnamefont{C.}~\bibnamefont{Mariani}}, \bibnamefont{and}
  \bibinfo{author}{\bibfnamefont{E.}~\bibnamefont{Vagnoni}},
  \bibinfo{journal}{Phys. Rev.} \textbf{\bibinfo{volume}{D92}},
  \bibinfo{pages}{091301} (\bibinfo{year}{2015}{\natexlab{b}}),
  \eprint{1507.08561}.

\bibitem[{\citenamefont{Cavanna et~al.}(2014)\citenamefont{Cavanna, Kordosky,
  Raaf, and Rebel}}]{Cavanna:2014iqa}
\bibinfo{author}{\bibfnamefont{F.}~\bibnamefont{Cavanna}},
  \bibinfo{author}{\bibfnamefont{M.}~\bibnamefont{Kordosky}},
  \bibinfo{author}{\bibfnamefont{J.}~\bibnamefont{Raaf}}, \bibnamefont{and}
  \bibinfo{author}{\bibfnamefont{B.}~\bibnamefont{Rebel}}
  (\bibinfo{collaboration}{LArIAT}) (\bibinfo{year}{2014}), \eprint{1406.5560}.

\bibitem[{\citenamefont{Agostino et~al.}(2014)}]{Agostino:2014qoa}
\bibinfo{author}{\bibfnamefont{L.}~\bibnamefont{Agostino}} \bibnamefont{et~al.}
  (\bibinfo{year}{2014}), \eprint{1409.4405}.

\bibitem[{\citenamefont{Kutter}(2015)}]{Kutter:2022751}
\bibinfo{author}{\bibfnamefont{T.}~\bibnamefont{Kutter}}, \bibinfo{type}{Tech.
  Rep.} \bibinfo{number}{CERN-SPSC-2015-020. SPSC-P-351},
  \bibinfo{institution}{CERN}, \bibinfo{address}{Geneva}
  (\bibinfo{year}{2015}), \bibinfo{note}{th. Kutter is acting as coordinator
  for the submitted proposal Spokespeople of the DUNE collaboration are Mark
  Thomson (Cambridge) Andre Rubbia (ETH Zuerich)},
  \urlprefix\url{https://cds.cern.ch/record/2022751}.

\bibitem[{\citenamefont{Berns et~al.}(2013)}]{Berns:2013usa}
\bibinfo{author}{\bibfnamefont{H.}~\bibnamefont{Berns}} \bibnamefont{et~al.}
  (\bibinfo{collaboration}{CAPTAIN}), in \emph{\bibinfo{booktitle}{{Community
  Summer Study 2013: Snowmass on the Mississippi (CSS2013) Minneapolis, MN,
  USA, July 29-August 6, 2013}}} (\bibinfo{year}{2013}), \eprint{1309.1740},
  \urlprefix\url{https://inspirehep.net/record/1253116/files/arXiv:1309.1740.pdf}.

\bibitem[{\citenamefont{Ankowski et~al.}(2010)}]{Ankowski:2008aa}
\bibinfo{author}{\bibfnamefont{A.}~\bibnamefont{Ankowski}} \bibnamefont{et~al.}
  (\bibinfo{collaboration}{ICARUS}), \bibinfo{journal}{Acta Phys. Polon.}
  \textbf{\bibinfo{volume}{B41}}, \bibinfo{pages}{103} (\bibinfo{year}{2010}),
  \eprint{0812.2373}.

\bibitem[{\citenamefont{Church}(2013)}]{Church:2013hea}
\bibinfo{author}{\bibfnamefont{E.~D.} \bibnamefont{Church}}
  (\bibinfo{year}{2013}), \eprint{1311.6774}.

\bibitem[{\citenamefont{Alion et~al.}(2016)}]{dunecdrconfigs}
\bibinfo{author}{\bibfnamefont{T.}~\bibnamefont{Alion}} \bibnamefont{et~al.}
  (\bibinfo{year}{2016}), \eprint{1606.09550}.

\bibitem[{\citenamefont{Andreopoulos et~al.}(2010)}]{Andreopoulos:2009rq}
\bibinfo{author}{\bibfnamefont{C.}~\bibnamefont{Andreopoulos}}
  \bibnamefont{et~al.}, \bibinfo{journal}{Nucl. Instrum. Meth.}
  \textbf{\bibinfo{volume}{A614}}, \bibinfo{pages}{87} (\bibinfo{year}{2010}),
  \eprint{0905.2517}.

\bibitem[{\citenamefont{Agostinelli et~al.}(2003)}]{Agostinelli:2002hh}
\bibinfo{author}{\bibfnamefont{S.}~\bibnamefont{Agostinelli}}
  \bibnamefont{et~al.} (\bibinfo{collaboration}{GEANT4}),
  \bibinfo{journal}{Nucl. Instrum. Meth.} \textbf{\bibinfo{volume}{A506}},
  \bibinfo{pages}{250} (\bibinfo{year}{2003}).

\bibitem[{\citenamefont{Yarba}(2012)}]{Yarba:2012ih}
\bibinfo{author}{\bibfnamefont{J.}~\bibnamefont{Yarba}}
  (\bibinfo{collaboration}{Geant4 Hadronic Group}), \bibinfo{journal}{J. Phys.
  Conf. Ser.} \textbf{\bibinfo{volume}{396}}, \bibinfo{pages}{022060}
  (\bibinfo{year}{2012}).

\bibitem[{\citenamefont{Szydagis et~al.}(2011)\citenamefont{Szydagis, Barry,
  Kazkaz, Mock, Stolp, Sweany, Tripathi, Uvarov, Walsh, and
  Woods}}]{Szydagis:2011tk}
\bibinfo{author}{\bibfnamefont{M.}~\bibnamefont{Szydagis}},
  \bibinfo{author}{\bibfnamefont{N.}~\bibnamefont{Barry}},
  \bibinfo{author}{\bibfnamefont{K.}~\bibnamefont{Kazkaz}},
  \bibinfo{author}{\bibfnamefont{J.}~\bibnamefont{Mock}},
  \bibinfo{author}{\bibfnamefont{D.}~\bibnamefont{Stolp}},
  \bibinfo{author}{\bibfnamefont{M.}~\bibnamefont{Sweany}},
  \bibinfo{author}{\bibfnamefont{M.}~\bibnamefont{Tripathi}},
  \bibinfo{author}{\bibfnamefont{S.}~\bibnamefont{Uvarov}},
  \bibinfo{author}{\bibfnamefont{N.}~\bibnamefont{Walsh}}, \bibnamefont{and}
  \bibinfo{author}{\bibfnamefont{M.}~\bibnamefont{Woods}},
  \bibinfo{journal}{JINST} \textbf{\bibinfo{volume}{6}},
  \bibinfo{pages}{P10002} (\bibinfo{year}{2011}), \eprint{1106.1613}.

\bibitem[{\citenamefont{Doke et~al.}(1988)\citenamefont{Doke, Crawford,
  Hitachi, Kikuchi, Lindstrom, Masuda, Shibamura, and Takahashi}}]{Doke:1988dp}
\bibinfo{author}{\bibfnamefont{T.}~\bibnamefont{Doke}},
  \bibinfo{author}{\bibfnamefont{H.~J.} \bibnamefont{Crawford}},
  \bibinfo{author}{\bibfnamefont{A.}~\bibnamefont{Hitachi}},
  \bibinfo{author}{\bibfnamefont{J.}~\bibnamefont{Kikuchi}},
  \bibinfo{author}{\bibfnamefont{P.~J.} \bibnamefont{Lindstrom}},
  \bibinfo{author}{\bibfnamefont{K.}~\bibnamefont{Masuda}},
  \bibinfo{author}{\bibfnamefont{E.}~\bibnamefont{Shibamura}},
  \bibnamefont{and}
  \bibinfo{author}{\bibfnamefont{T.}~\bibnamefont{Takahashi}},
  \bibinfo{journal}{Nucl. Instrum. Meth.} \textbf{\bibinfo{volume}{A269}},
  \bibinfo{pages}{291} (\bibinfo{year}{1988}).

\bibitem[{\citenamefont{Miyajima et~al.}(1974)\citenamefont{Miyajima,
  Takahashi, Konno, Hamada, Kubota, Shibamura, and Doke}}]{Miyajima:1974zz}
\bibinfo{author}{\bibfnamefont{M.}~\bibnamefont{Miyajima}},
  \bibinfo{author}{\bibfnamefont{T.}~\bibnamefont{Takahashi}},
  \bibinfo{author}{\bibfnamefont{S.}~\bibnamefont{Konno}},
  \bibinfo{author}{\bibfnamefont{T.}~\bibnamefont{Hamada}},
  \bibinfo{author}{\bibfnamefont{S.}~\bibnamefont{Kubota}},
  \bibinfo{author}{\bibfnamefont{H.}~\bibnamefont{Shibamura}},
  \bibnamefont{and} \bibinfo{author}{\bibfnamefont{T.}~\bibnamefont{Doke}},
  \bibinfo{journal}{Phys. Rev.} \textbf{\bibinfo{volume}{A9}},
  \bibinfo{pages}{1438} (\bibinfo{year}{1974}).

\bibitem[{\citenamefont{Acciarri et~al.}(2016)}]{Acciarri:2016ooe}
\bibinfo{author}{\bibfnamefont{R.}~\bibnamefont{Acciarri}} \bibnamefont{et~al.}
  (\bibinfo{collaboration}{DUNE}) (\bibinfo{year}{2016}), \eprint{1601.02984}.

\bibitem[{\citenamefont{Adams et~al.}(2013)}]{Adams:2013qkq}
\bibinfo{author}{\bibfnamefont{C.}~\bibnamefont{Adams}} \bibnamefont{et~al.}
  (\bibinfo{collaboration}{LBNE}) (\bibinfo{year}{2013}), \eprint{1307.7335}.

\bibitem[{\citenamefont{Huber et~al.}(2005)\citenamefont{Huber, Lindner, and
  Winter}}]{Huber:2004ka}
\bibinfo{author}{\bibfnamefont{P.}~\bibnamefont{Huber}},
  \bibinfo{author}{\bibfnamefont{M.}~\bibnamefont{Lindner}}, \bibnamefont{and}
  \bibinfo{author}{\bibfnamefont{W.}~\bibnamefont{Winter}},
  \bibinfo{journal}{Comput. Phys. Commun.} \textbf{\bibinfo{volume}{167}},
  \bibinfo{pages}{195} (\bibinfo{year}{2005}), \eprint{hep-ph/0407333}.

\bibitem[{\citenamefont{Huber et~al.}(2007)\citenamefont{Huber, Kopp, Lindner,
  Rolinec, and Winter}}]{Huber:2007ji}
\bibinfo{author}{\bibfnamefont{P.}~\bibnamefont{Huber}},
  \bibinfo{author}{\bibfnamefont{J.}~\bibnamefont{Kopp}},
  \bibinfo{author}{\bibfnamefont{M.}~\bibnamefont{Lindner}},
  \bibinfo{author}{\bibfnamefont{M.}~\bibnamefont{Rolinec}}, \bibnamefont{and}
  \bibinfo{author}{\bibfnamefont{W.}~\bibnamefont{Winter}},
  \bibinfo{journal}{Comput. Phys. Commun.} \textbf{\bibinfo{volume}{177}},
  \bibinfo{pages}{432} (\bibinfo{year}{2007}), \eprint{hep-ph/0701187}.

\bibitem[{\citenamefont{Dziewonski and Anderson}(1981)}]{Dziewonski:1981xy}
\bibinfo{author}{\bibfnamefont{A.~M.} \bibnamefont{Dziewonski}}
  \bibnamefont{and} \bibinfo{author}{\bibfnamefont{D.~L.}
  \bibnamefont{Anderson}}, \bibinfo{journal}{Phys. Earth Planet. Interiors}
  \textbf{\bibinfo{volume}{25}}, \bibinfo{pages}{297} (\bibinfo{year}{1981}).

\bibitem[{\citenamefont{Stacey}(1977)}]{stacey}
\bibinfo{author}{\bibfnamefont{F.}~\bibnamefont{Stacey}},
  \emph{\bibinfo{title}{{Physics of the earth}}} (\bibinfo{publisher}{Wiley},
  \bibinfo{year}{1977}), \bibinfo{edition}{2nd} ed.

\bibitem[{\citenamefont{Cervera et~al.}(2000)\citenamefont{Cervera, Donini,
  Gavela, Gomez~Cadenas, Hernandez et~al.}}]{Cervera:2000kp}
\bibinfo{author}{\bibfnamefont{A.}~\bibnamefont{Cervera}},
  \bibinfo{author}{\bibfnamefont{A.}~\bibnamefont{Donini}},
  \bibinfo{author}{\bibfnamefont{M.}~\bibnamefont{Gavela}},
  \bibinfo{author}{\bibfnamefont{J.}~\bibnamefont{Gomez~Cadenas}},
  \bibinfo{author}{\bibfnamefont{P.}~\bibnamefont{Hernandez}},
  \bibnamefont{et~al.}, \bibinfo{journal}{Nucl.Phys.}
  \textbf{\bibinfo{volume}{B579}}, \bibinfo{pages}{17} (\bibinfo{year}{2000}),
  \eprint{hep-ph/0002108}.

\bibitem[{\citenamefont{Freund}(2001)}]{Freund:2001pn}
\bibinfo{author}{\bibfnamefont{M.}~\bibnamefont{Freund}},
  \bibinfo{journal}{Phys. Rev.} \textbf{\bibinfo{volume}{D64}},
  \bibinfo{pages}{053003} (\bibinfo{year}{2001}), \eprint{hep-ph/0103300}.

\bibitem[{\citenamefont{Diwan et~al.}(2003)}]{Diwan:2003bp}
\bibinfo{author}{\bibfnamefont{M.~V.} \bibnamefont{Diwan}}
  \bibnamefont{et~al.}, \bibinfo{journal}{Phys. Rev.}
  \textbf{\bibinfo{volume}{D68}}, \bibinfo{pages}{012002}
  (\bibinfo{year}{2003}), \eprint{hep-ph/0303081}.

\bibitem[{\citenamefont{Barger et~al.}(2006)\citenamefont{Barger, Dierckxsens,
  Diwan, Huber, Lewis, Marfatia, and Viren}}]{Barger:2006vy}
\bibinfo{author}{\bibfnamefont{V.}~\bibnamefont{Barger}},
  \bibinfo{author}{\bibfnamefont{M.}~\bibnamefont{Dierckxsens}},
  \bibinfo{author}{\bibfnamefont{M.}~\bibnamefont{Diwan}},
  \bibinfo{author}{\bibfnamefont{P.}~\bibnamefont{Huber}},
  \bibinfo{author}{\bibfnamefont{C.}~\bibnamefont{Lewis}},
  \bibinfo{author}{\bibfnamefont{D.}~\bibnamefont{Marfatia}}, \bibnamefont{and}
  \bibinfo{author}{\bibfnamefont{B.}~\bibnamefont{Viren}},
  \bibinfo{journal}{Phys. Rev.} \textbf{\bibinfo{volume}{D74}},
  \bibinfo{pages}{073004} (\bibinfo{year}{2006}), \eprint{hep-ph/0607177}.

\bibitem[{\citenamefont{Barger et~al.}(2007)\citenamefont{Barger, Bishai,
  Bogert, Bromberg, Curioni et~al.}}]{Barger:2007yw}
\bibinfo{author}{\bibfnamefont{V.}~\bibnamefont{Barger}},
  \bibinfo{author}{\bibfnamefont{M.}~\bibnamefont{Bishai}},
  \bibinfo{author}{\bibfnamefont{D.}~\bibnamefont{Bogert}},
  \bibinfo{author}{\bibfnamefont{C.}~\bibnamefont{Bromberg}},
  \bibinfo{author}{\bibfnamefont{A.}~\bibnamefont{Curioni}},
  \bibnamefont{et~al.} (\bibinfo{year}{2007}), \eprint{0705.4396}.

\bibitem[{\citenamefont{Huber and Kopp}(2011)}]{Huber:2010dx}
\bibinfo{author}{\bibfnamefont{P.}~\bibnamefont{Huber}} \bibnamefont{and}
  \bibinfo{author}{\bibfnamefont{J.}~\bibnamefont{Kopp}},
  \bibinfo{journal}{JHEP} \textbf{\bibinfo{volume}{03}}, \bibinfo{pages}{013}
  (\bibinfo{year}{2011}), \bibinfo{note}{[Erratum: JHEP05,024(2011)]},
  \eprint{1010.3706}.

\bibitem[{\citenamefont{Ghosh et~al.}(2014)\citenamefont{Ghosh, Goswami, and
  Raut}}]{Ghosh:2014rna}
\bibinfo{author}{\bibfnamefont{M.}~\bibnamefont{Ghosh}},
  \bibinfo{author}{\bibfnamefont{S.}~\bibnamefont{Goswami}}, \bibnamefont{and}
  \bibinfo{author}{\bibfnamefont{S.~K.} \bibnamefont{Raut}}
  (\bibinfo{year}{2014}), \eprint{1412.1744}.

\bibitem[{\citenamefont{Coloma and Fernandez-Martinez}(2012)}]{Coloma:2011pg}
\bibinfo{author}{\bibfnamefont{P.}~\bibnamefont{Coloma}} \bibnamefont{and}
  \bibinfo{author}{\bibfnamefont{E.}~\bibnamefont{Fernandez-Martinez}},
  \bibinfo{journal}{JHEP} \textbf{\bibinfo{volume}{1204}}, \bibinfo{pages}{089}
  (\bibinfo{year}{2012}), \eprint{1110.4583}.

\bibitem[{\citenamefont{Baussan et~al.}(2014)}]{Baussan:2013zcy}
\bibinfo{author}{\bibfnamefont{E.}~\bibnamefont{Baussan}} \bibnamefont{et~al.}
  (\bibinfo{collaboration}{ESSnuSB Collaboration}), \bibinfo{journal}{Nuclear
  Physics B} \textbf{\bibinfo{volume}{885}}, \bibinfo{pages}{127 }
  (\bibinfo{year}{2014}), \eprint{1309.7022}.

\bibitem[{\citenamefont{Agarwalla et~al.}(2014)\citenamefont{Agarwalla,
  Choubey, and Prakash}}]{Agarwalla:2014tpa}
\bibinfo{author}{\bibfnamefont{S.~K.} \bibnamefont{Agarwalla}},
  \bibinfo{author}{\bibfnamefont{S.}~\bibnamefont{Choubey}}, \bibnamefont{and}
  \bibinfo{author}{\bibfnamefont{S.}~\bibnamefont{Prakash}},
  \bibinfo{journal}{JHEP} \textbf{\bibinfo{volume}{12}}, \bibinfo{pages}{020}
  (\bibinfo{year}{2014}), \eprint{1406.2219}.

\bibitem[{\citenamefont{Blennow et~al.}(2014)\citenamefont{Blennow, Coloma,
  Huber, and Schwetz}}]{Blennow:2013oma}
\bibinfo{author}{\bibfnamefont{M.}~\bibnamefont{Blennow}},
  \bibinfo{author}{\bibfnamefont{P.}~\bibnamefont{Coloma}},
  \bibinfo{author}{\bibfnamefont{P.}~\bibnamefont{Huber}}, \bibnamefont{and}
  \bibinfo{author}{\bibfnamefont{T.}~\bibnamefont{Schwetz}},
  \bibinfo{journal}{JHEP} \textbf{\bibinfo{volume}{1403}}, \bibinfo{pages}{028}
  (\bibinfo{year}{2014}), \eprint{1311.1822}.

\bibitem[{\citenamefont{Forero et~al.}(2014)\citenamefont{Forero, Tortola, and
  Valle}}]{Forero:2014bxa}
\bibinfo{author}{\bibfnamefont{D.~V.} \bibnamefont{Forero}},
  \bibinfo{author}{\bibfnamefont{M.}~\bibnamefont{Tortola}}, \bibnamefont{and}
  \bibinfo{author}{\bibfnamefont{J.~W.~F.} \bibnamefont{Valle}},
  \bibinfo{journal}{Phys. Rev.} \textbf{\bibinfo{volume}{D90}},
  \bibinfo{pages}{093006} (\bibinfo{year}{2014}), \eprint{1405.7540}.

\bibitem[{\citenamefont{Capozzi et~al.}(2014)\citenamefont{Capozzi, Fogli,
  Lisi, Marrone, Montanino, and Palazzo}}]{Capozzi:2013csa}
\bibinfo{author}{\bibfnamefont{F.}~\bibnamefont{Capozzi}},
  \bibinfo{author}{\bibfnamefont{G.~L.} \bibnamefont{Fogli}},
  \bibinfo{author}{\bibfnamefont{E.}~\bibnamefont{Lisi}},
  \bibinfo{author}{\bibfnamefont{A.}~\bibnamefont{Marrone}},
  \bibinfo{author}{\bibfnamefont{D.}~\bibnamefont{Montanino}},
  \bibnamefont{and} \bibinfo{author}{\bibfnamefont{A.}~\bibnamefont{Palazzo}},
  \bibinfo{journal}{Phys. Rev.} \textbf{\bibinfo{volume}{D89}},
  \bibinfo{pages}{093018} (\bibinfo{year}{2014}), \eprint{1312.2878}.

\bibitem[{\citenamefont{Schwetz}(2007)}]{Schwetz:2006md}
\bibinfo{author}{\bibfnamefont{T.}~\bibnamefont{Schwetz}},
  \bibinfo{journal}{Phys.Lett.} \textbf{\bibinfo{volume}{B648}},
  \bibinfo{pages}{54} (\bibinfo{year}{2007}), \eprint{hep-ph/0612223}.

\bibitem[{\citenamefont{Blennow et~al.}(2015)\citenamefont{Blennow, Coloma, and
  Fernandez-Martinez}}]{Blennow:2014sja}
\bibinfo{author}{\bibfnamefont{M.}~\bibnamefont{Blennow}},
  \bibinfo{author}{\bibfnamefont{P.}~\bibnamefont{Coloma}}, \bibnamefont{and}
  \bibinfo{author}{\bibfnamefont{E.}~\bibnamefont{Fernandez-Martinez}},
  \bibinfo{journal}{JHEP} \textbf{\bibinfo{volume}{03}}, \bibinfo{pages}{005}
  (\bibinfo{year}{2015}), \eprint{1407.3274}.

\bibitem[{\citenamefont{Elevant and Schwetz}(2015)}]{Elevant:2015ska}
\bibinfo{author}{\bibfnamefont{J.}~\bibnamefont{Elevant}} \bibnamefont{and}
  \bibinfo{author}{\bibfnamefont{T.}~\bibnamefont{Schwetz}},
  \bibinfo{journal}{JHEP} \textbf{\bibinfo{volume}{09}}, \bibinfo{pages}{016}
  (\bibinfo{year}{2015}), \eprint{1506.07685}.

\bibitem[{\citenamefont{Coloma et~al.}(2012)\citenamefont{Coloma, Donini,
  Fernandez-Martinez, and Hernandez}}]{Coloma:2012wq}
\bibinfo{author}{\bibfnamefont{P.}~\bibnamefont{Coloma}},
  \bibinfo{author}{\bibfnamefont{A.}~\bibnamefont{Donini}},
  \bibinfo{author}{\bibfnamefont{E.}~\bibnamefont{Fernandez-Martinez}},
  \bibnamefont{and}
  \bibinfo{author}{\bibfnamefont{P.}~\bibnamefont{Hernandez}},
  \bibinfo{journal}{JHEP} \textbf{\bibinfo{volume}{06}}, \bibinfo{pages}{073}
  (\bibinfo{year}{2012}), \eprint{1203.5651}.

\end{thebibliography}
